\documentclass[oldversion]{aa}  

\usepackage{graphicx}
\usepackage{txfonts}
\usepackage{longtable}

\usepackage{natbib}
\usepackage{times}
\bibpunct{(}{)}{;}{a}{}{,}

\def\0{\phantom0}

\def\zl{$z_{\rm lens}$}

\def\ho{H$_{\rm 0}$}

\def\kmsmpc{km s$^{-1}$ Mpc$^{-1}$}

\def\obj{WFI~J2033--4723}

\begin{document}

\title{COSMOGRAIL: the COSmological  MOnitoring of \\ \vspace*{1mm}
  GRAvItational Lenses\thanks{Based on observations obtained with 
  the 1.2m EULER Swiss Telescope,
  the 1.3m Small and Moderate Aperture Research Telescope System (SMARTS)
  which is operated by the SMARTS Consortium, and
  the NASA/ESA {\it Hubble Space Telescope} as part of program HST-GO-9744
  of the Space Telescope Science Institute, which is operated by the
  Association of Universities for Research in Astronomy, Inc., under
  NASA contract NAS~5-26555.  }
  } 

\subtitle{VII. Time delays and the Hubble constant from \obj}

\titlerunning{COSMOGRAIL~VII: time delays and \ho\ from \obj}

\author{C.~Vuissoz\inst{1} \and F.~Courbin\inst{1} 
  \and D.~Sluse\inst{1} \and G.~Meylan\inst{1}
  \and V.~Chantry\inst{2}\thanks{Research Fellow, Belgian National Fund for Scientific Research (FNRS)}
  \and E.~Eulaers\inst{2} 
  \and C.~Morgan\inst{3,4} \and M.E.~Eyler\inst{4} \and C.S.~Kochanek\inst{3}
  \and J.~Coles\inst{5} \and P.~Saha\inst{5} 
  \and P.~Magain\inst{2}
  \and E.E.~Falco\inst{6}
  } 

\institute{Laboratoire d'Astrophysique, Ecole Polytechnique
     F\'ed\'erale de Lausanne (EPFL), Observatoire de Sauverny,
     CH-1290 Versoix, Switzerland 
     \and
     Institut d'Astrophysique et de G\'eophysique, Universit\'e de
     Li\`ege, All\'ee du 6 ao\^ut 17, Sart-Tilman, B\^at. B5C, 4000
     Li\`ege, Belgium
     \and
     Department of Astronomy and the Center for Cosmology and Astroparticle Physics, 
     The Ohio State University, Columbus, OH 43210, USA
     \and
     Department of Physics, United States Naval Academy, 572C Holloway Road, Annapolis MD 21402, USA
     \and
     Institute of Theoretical Physics, University of Z\"urich, 
     Winterthurerstrasse 190, 8057 Z\"urich, Switzerland
     \and
     Harvard-Smithsonian Center for Astrophysics, 60 Garden Street, Cambridge MA 02138, USA
     } 

\date{Received 31 March 2008 / Accepted 7 July 2008}

\abstract{Gravitationally lensed  quasars can be used to  map the mass
  distribution in lensing galaxies and to estimate the Hubble constant
  \ho\ by measuring  the time delays between the  quasar images.  Here
  we  report the  measurement of  two independent  time delays  in the
  quadruply  imaged  quasar \obj\  ($z=1.66$).   Our  data consist  of
  $R$-band  images  obtained with  the  Swiss  1.2\,m EULER  telescope
  located at La Silla and  with the 1.3\,m SMARTS telescope located at
  Cerro Tololo. The light  curves have 218 independent epochs spanning
  3 full years  of monitoring between March 2004 and  May 2007, with a
  mean temporal sampling of one observation every 4th day.  We measure
  the  time delays  using three  different techniques,  and  we obtain
  $\Delta t_{B-A} =  35.5 \pm 1.4$~days (3.8\%) and  $\Delta t_{B-C} =
  62.6~^{+\,4.1}_{-\,2.3}~\  \hbox{days} ~\ (^{+\,6.5\%}_{-\,3.7\%})$, 
  where $A$ is a composite of the close, merging image pair.   
  After correcting for the time  delays, we find
  $R$-band flux ratios of  $F_{A}/F_{B} = 2.88 \pm 0.04$, $F_{A}/F_{C}
  = 3.38 \pm  0.06$, and $F_{A1}/F_{A2} = 1.37 \pm  0.05$  with no
    evidence for  microlensing variability over a time  scale of three
    years. However, these flux ratios do not agree with those measured
    in  the  quasar  emission   lines,  suggesting  that  longer  term
    microlensing is  present.  Our estimate  of \ho\ agrees  with the
  concordance  value: non-parametric  modeling of  the  lensing galaxy
  predicts  \ho\  =  67   $^{+13}_{-10}$  \kmsmpc,  while  the  Single
  Isothermal Sphere model yields  \ho\ = 63 $^{+7}_{-3}$ \kmsmpc (68\%
  confidence level).  More  complex lens models using a composite
    de Vaucouleurs plus  NFW galaxy mass profile show  twisting of the
    mass isocontours  in the lensing galaxy, as  do the non-parametric
    models.  As all models  also require a significant external shear,
    this suggests that  the lens is a member of  the group of galaxies
    seen in field of view of \obj. }


\maketitle

\begin{figure*}[ht!]
  \begin{center}
    \includegraphics[width=12cm]{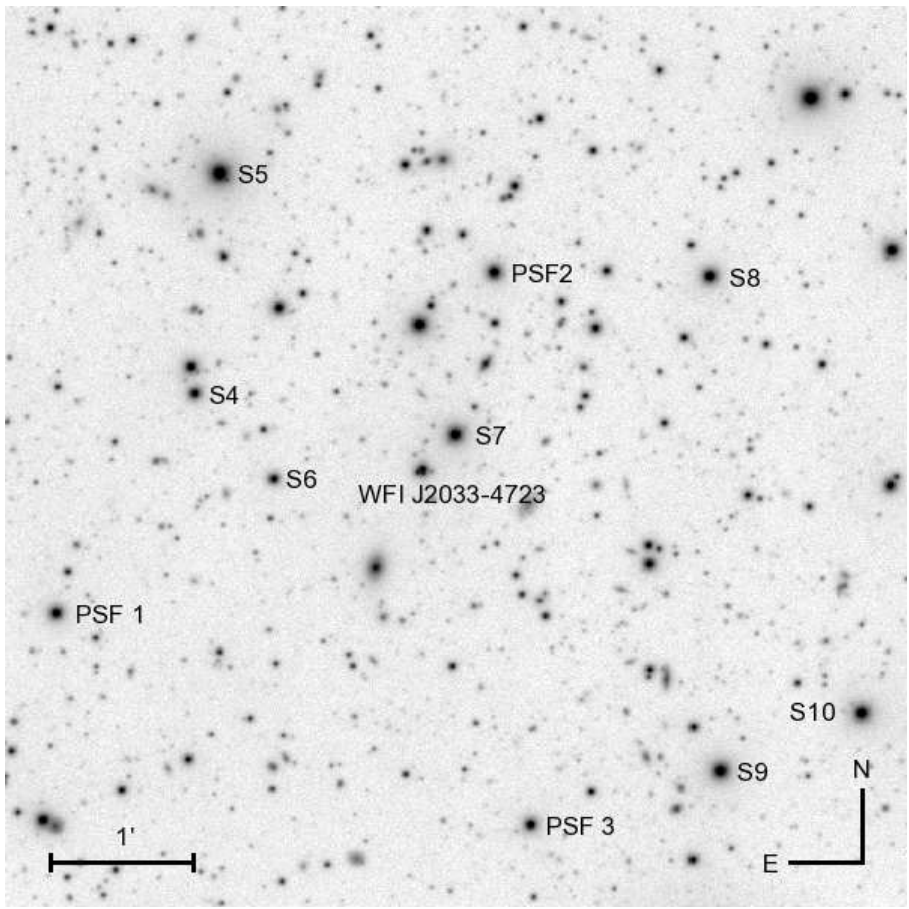}
    \caption{The 6.3\arcmin\ $\times$ 6.3\arcmin\ field of view around
      \obj.  This  image is the central  part of a  combination of 418
      $R$-band frames  obtained with the  1.2m EULER Telescope  with a
      total exposure time of 48~hours and a mean seeing of 1.3\arcsec.
      The three stars PSF1--3 used  to model the Point Spread Function
      (PSF) and the 7 reference  stars S4--10 used for frame alignment
      and flux calibration are indicated.}
    \label{field}
  \end{center}
\end{figure*}

\section{Introduction}

When a quasar is gravitationally lensed and we observe multiple images
of  the source  there are  light travel  time differences  between the
images.   Any  intrinsic  variation  of  the  quasar  is  observed  at
different times in each image with a measurable ``time delay'' between
them.   \citet{refsdal1964} first  showed that  time delays  provide a
means of determining the Hubble constant \ho\ independent of any local
distance calibrator,  provided a  mass model can  be inferred  for the
lensing  galaxy.  Conversely,  one can  also assume  \ho\ in  order to
study the distribution of the total mass in the lensing galaxy.

During the  past 25 years, time  delays have been measured  in only 17
systems,   at    various   accuracy   levels    \citep[see][   for   a
  review]{oguri2007}.   As  the error  in  the  time delay  propagates
directly  into  \ho,   it  is  important  to  make   it  as  small  as
possible. Unfortunately, most  existing time delays have uncertainties
of the order of 10\%  that are comparable to the current uncertainties
in \ho.  This  uncertainty can be reduced by  increasing the sample of
lenses  with known  time  delays, and  by  simultaneously fitting  all
lenses in  the sample with  a common value for  \ho\ \citep{saha2006a,
  saha2006b, coles2008}.

COSMOGRAIL is an optical monitoring campaign that aims to measure time
delays  for  a  large  number  of gravitationally  lensed  quasars  to
accuracies  of a  few percent  using  a network  of 1-  and 2-m  class
telescopes.  The first result of  this campaign was the measurement of
the  time delay  in the  doubly  imaged quasar  SDSS~J1650+4251 to  an
accuracy   of  3.8\%   based  on   two  observing   seasons   of  data
\citep{vuissoz2007}. COSMOGRAIL complements  a second monitoring group
whose   most   recent   results   are   a   delay   for   HE~1104-1805
\citep{poindexter2007}.    In  this   paper   we  present   time-delay
measurements for the quadruply imaged  quasar \obj\ based on merging 3
years of optical monitoring data  from the two groups.  In a companion
effort, \citet{morgan2008} analyzed the  merged data for the two-image
lens  QJ0158--4325, succeeding  in measuring  the size  of  the source
accretion disk  but failing to  measure a time  delay due to  the high
amplitude of the microlensing variability in this system.

\obj\     (20$^{\rm h}$33$^{\rm m}$42\fs08,
--47\degr23\arcmin43\farcs0;     J2000.0)     was    discovered     by
\citet{morgan2004} and consists of 4 images of a $z =1.66$ quasar with
a maximum separation of 2.5\arcsec.  The lens galaxy was identified by
\citet{morgan2004} and  has a spectroscopic redshift of  \zl\ = $0.661
\pm 0.001$ \citep{eigenbrod2006b}.  The  lens appears to be the member
of a  group, with at  least 6 galaxies  within 20\arcsec\ of  the lens
\citep{morgan2004}, and  we will have to account  for this environment
in any lens model.

We   describe    the   monitoring    data   and   its    analysis   in
Sect.~\ref{monitoring}  In  Sect.~\ref{HST}  we  present  the  near-IR
\emph{Hubble  Space  Telescope} (HST)  observations  that  we used  to
obtain  accurate differential  astrometry of  the lens  components and
surface photometry of the lens galaxy.  We estimate the time delays in
Sect.~\ref{timedelays}    and     model    them    using    parametric
(Sect.~\ref{param_models})              and             non-parametric
(Sect.~\ref{saha_models}) models for the mass distribution of the lens
galaxy.  We summarize our results in Sec.~\ref{conclusion}.

\section{Photometric monitoring}
\label{monitoring}

Our 3-year photometric monitoring of  \obj\ was carried out from March
2004 to May 2007 with the 1.2\,m EULER telescope and the 1.3\,m SMARTS
telescope  located  in  Chile  at   La  Silla  and  the  Cerro  Tololo
Interamerican Observatory  (CTIO), respectively.  \obj\  was monitored
from  both sites  for three  years during  its visibility  window from
early March to mid-December.

The  1.2m EULER  telescope  is equipped  with  a 2048$\times$2048  CCD
camera which  has 0.344\arcsec\ pixels  and produces an image  with an
11\arcmin\ field of view.  The mean  sampling of the EULER data is one
epoch every five days, where each epoch consists of five dithered 360s
images taken with an R-band filter.   The worst gaps due to weather or
technical problems are 2--3 weeks.  The EULER data set consists of 141
epochs  of data  obtained between  May 2004  and May  2007.  The
image quality  varies between 0.9\arcsec\ and 2.0\arcsec\  FWHM over 3
years, with a median of 1.4\arcsec.

The  1.3m SMARTS  telescope  is equipped  with  the dual-beam  ANDICAM
\citep{depoy2003}  camera.  Here  we use  only the  optical channel
which        has        0.369\arcsec\        pixels       and        a
6.5\arcmin$\times$6.3\arcmin\ field of view.  The mean sampling of the
SMARTS data is  one epoch every eight days,  with three 300s exposures
at each  epoch.  The  SMARTS data  set consists of  77 epochs  of data
obtained  between March  2004 and  December 2006.   The seeing  on the
images  varies between 0.5\arcsec\  and 2.0\arcsec,  with a  median of
1.4\arcsec.

The combined data set consists  of 218 observing epochs comprising 956
images covering  the common field  of view shown  in Fig.~\ref{field}.
The  average   temporal  sampling  when  \obj\  was   visible  is  one
observation every 4 days over a period of three years, one of the best
sets of monitoring data available for a lensed quasar.

\begin{figure}[t!]
  \begin{center}
    \includegraphics[width=8.0cm]{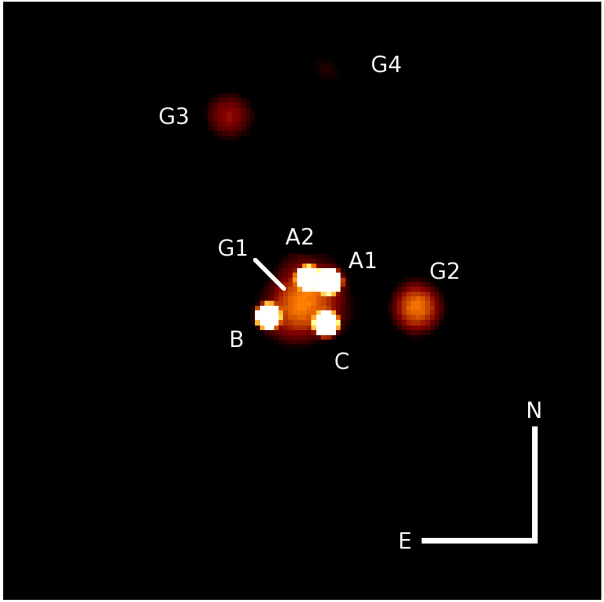}
    \caption{Result  of  the  simultaneous  deconvolution of  the  956
      $R$-band images (EULER+SMARTS) of  \obj.  The pixel size of this
      image  is half  the  pixel  size of  the  EULER detector,  i.e.,
      0.172\arcsec,    and     the    resolution    is     2    pixels
      Full-Width-Half-Maximum, i.e., 0.344\arcsec.   The field of view
      is 22\arcsec\ on a side.  Two galaxies $G2$ and $G3$ are seen to
      the West  and North  of the main  lensing galaxy $G1$.   $G3$ is
      part of  a group included in  the lens modeling,  while $G1$ and
      $G2$ are modeled individually (see Sect.~\ref{param_models}).}
    \label{deconvim}
  \end{center}
\end{figure}

The EULER data  are reduced using the automated  pipeline described in
\citet{vuissoz2007}  and the  SMARTS  data with  the SMARTS  pipeline,
using  standard methods.   The  reduced frames  are  then aligned  and
interpolated  to a  common  reference frame,  one  of the  best-seeing
(1\arcsec) EULER images,  taken on the night of 5  April 2006.  The 10
stars  (PSF1--3 and  S4--10)  shown in  Fig.~\ref{field}  are used  to
determine  the  geometric transformation  needed  for  each EULER  and
SMARTS  image  to  match  the  reference  frame.   The  transformation
includes  image parity,  rotation, shifting  and rescaling.   These 10
stars are  also used to determine  the photometric zero  point of each
image  relative to  the reference  image. After  interpolation, cosmic
rays are removed using the L.A.Cosmic algorithm \citep{vanDokkum2001}.
We check that no data pixels are mistakenly removed by this process.

The  light  curves  of  the  quasars are  measured  by  simultaneously
deconvolving    all    the   images    using    the   MCS    algorithm
\citep{magain1998}.  This method has already been successfully applied
to    the    monitoring     data    of    several    lensed    quasars
\citep[e.g.][]{vuissoz2007, hjorth2002,  burud2002a, burud2002b}.  The
deconvolved images  have a pixel scale of  0.172\arcsec\ (one-half the
pixel scale of the EULER data)  and are constructed to have a Gaussian
PSF with  a 2  pixel (0.344\arcsec) FWHM.   The Point  Spread Function
(PSF) for each  of the 956 images is constructed  from the three stars
PSF1--3 (see Fig.~\ref{field}).  During the deconvolution process, the
relative positions of  the quasar images are held  fixed to the values
derived from  fitting the HST  images in Sect.~\ref{HST},  while their
fluxes are  allowed to vary from  frame to frame.   The flux, position
and shape of  the lensing galaxy are the same for  all frames, but the
values vary freely as part of the fit.

\begin{figure*}[ht]
  \begin{center}
    \includegraphics[width=14.0cm]{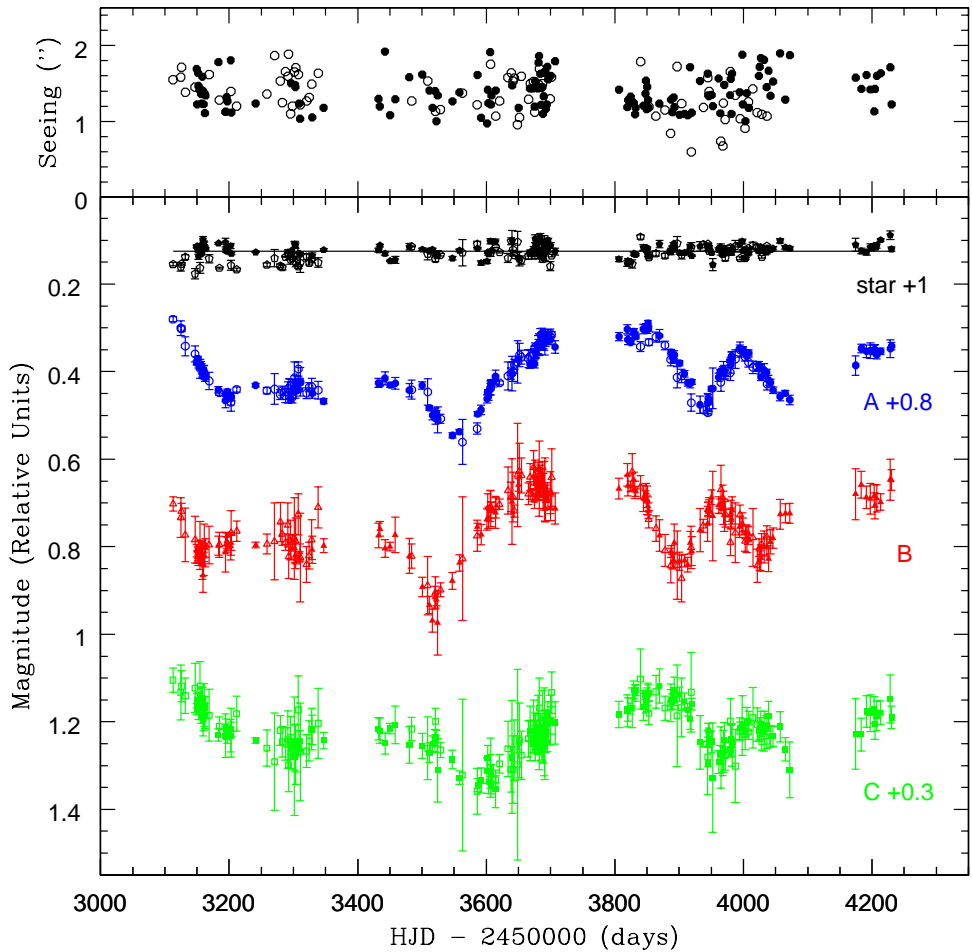}
    \caption{Our $R$-band  light curves obtained for \obj\  as well as
      for   the  reference  star   S6  (see   Fig.~\ref{field}).   The
      magnitudes  are given  in  relative units.   The filled  symbols
      correspond  to  the EULER  observations  while  the SMARTS  data
      points are  marked by open symbols.  Components  $A_1$ and $A_2$
      were summed into one single component $A$.  The curves have been
      shifted  in  magnitude  for  visibility,  as  indicated  in  the
      figure.}
    \label{lightcurves}
  \end{center}
\end{figure*}

Fig.~\ref{deconvim} shows  an example of  a deconvolved image.   It is
clear  that  we will  have  no  difficulty  estimating the  fluxes  of
components  $B$  and  $C$  separately.  Components  $A_1$  and  $A_2$,
however, are separated  by only 0.724\arcsec, which is  only twice the
resolution  of our  deconvolved images,  and remain  partially blended
after deconvolution.   Since the delay between these  images should be
very small, we will sum the fluxes of the two images and consider only
the light curve of the total flux $A=A_1+A_2$.  The resulting $R$-band
light curves are displayed in Fig.~\ref{lightcurves}.

We also display in  Fig.~\ref{lightcurves} the deconvolved light curve
of the isolated  star $S6$, which has roughly the  same color as \obj.
Each point is  the mean of the  images taken at a given  epoch and the
error  bar is the  1$\sigma$ standard  error of  the mean.   The light
curve  is  flat,  with  a  standard  deviation over  the  3  years  of
monitoring  of  $\sigma_{\rm{tot}} =  0.010$\,mag  about its  average,
which   is   roughly  consistent   with   the   mean   error  bar   of
$\sigma_{\rm{mean}} = 0.006$\,mag of the individual epochs.

The dispersion of  the points in the star's  light curve reflects both
statistical  errors   and  systematic  errors   from  the  photometric
calibrations and the construction of  the PSF.  To the extent that all
the  light curves  suffer  from  the same  systematic  errors, we  can
correct the quasar's light curves  by subtracting the residuals of the
star's light curve from each  of them.  We then define the uncertainty
in a quasar's  light curve as the quadrature  sum of the uncertainties
in  the two  light curves.   This procedure  will increase  the photon
noise but should minimize the systematic errors.

\section{HST Near-IR Imaging}
\label{HST}

\begin{figure*}[ht]
  \centering
  \includegraphics[width=6.0cm]{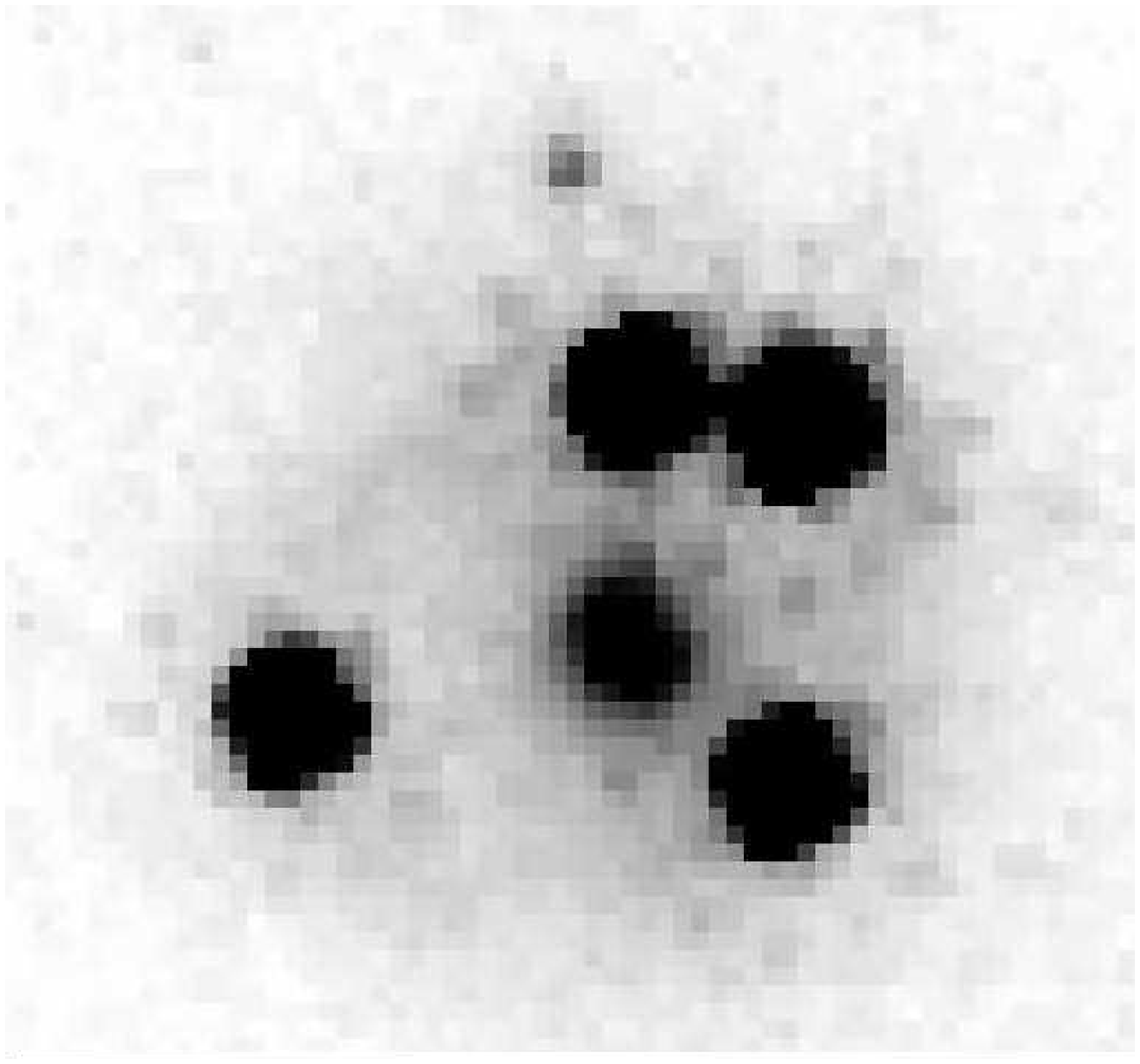}
  \includegraphics[width=6.0cm]{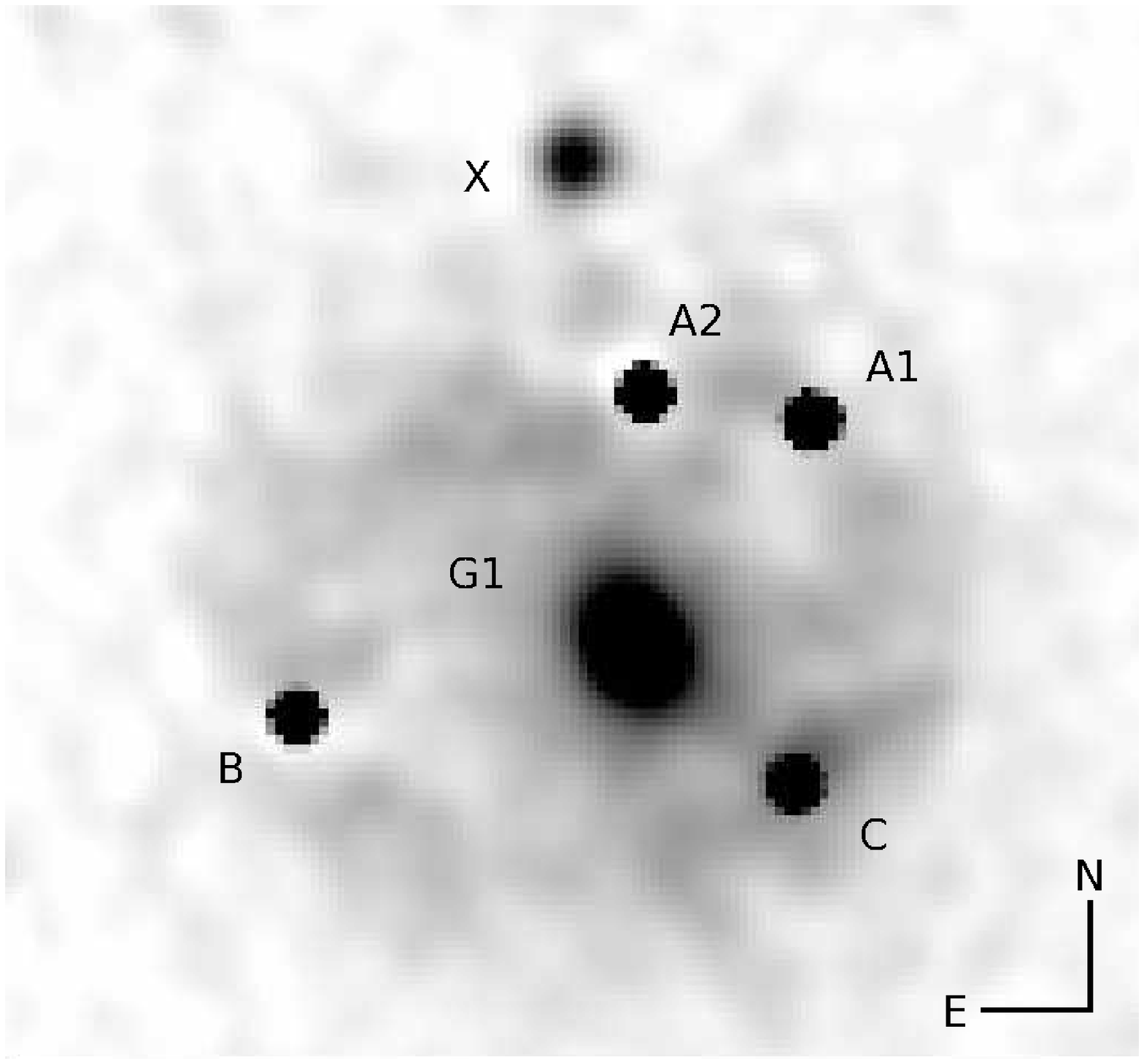}
  \includegraphics[width=6.0cm]{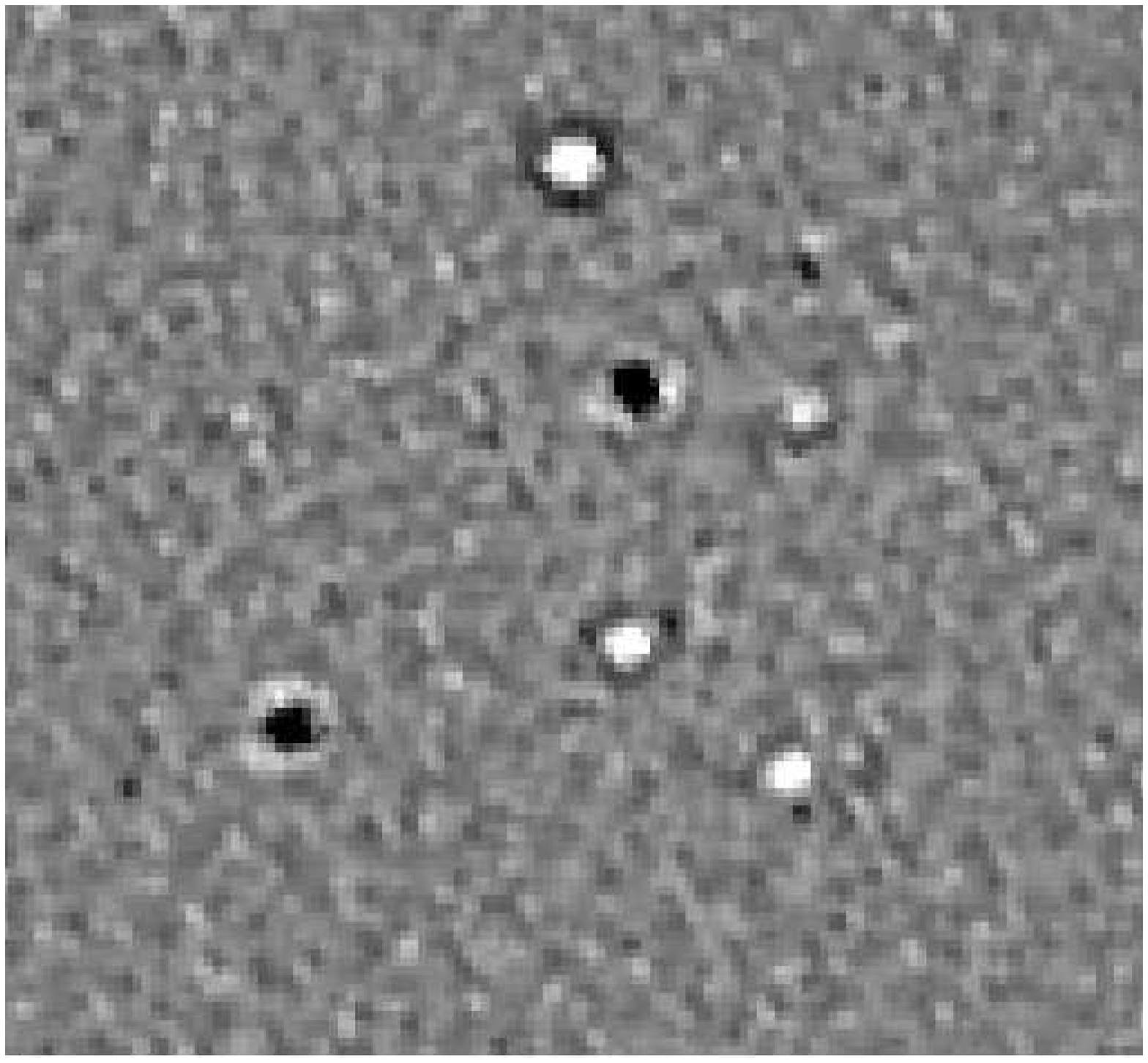}
  \setlength{\unitlength}{1cm}
  \caption{   {\it   Left}:  Deep   NICMOS2   image,   taken  in   the
    F160W-band. This image  is a combination of 4  frames, for a total
    exposure time  of 46~minutes.  North is  up and East  to the left.
    The  field  of  view  is  4\arcsec\  on  a  side.   {\it  Middle}:
    Simultaneous  deconvolution of the  individual NICMOS  images (see
    text),  using the  MCS deconvolution  algorithm. The  PSF  in this
    image    is    an    analytical    Gaussian    with    2    pixels
    Full-Width-at-Half-Maximum   (FWHM),  i.e.,   the   resolution  is
    0.075\arcsec.  The  pixel size is  0.035\arcsec, i.e., oversampled
    by  a factor of  two compared  to the  original pixel  size.  {\it
      Right}: Residual  map of the deconvolution, with  the cut levels
    set  to $\pm  5\sigma$.  Only  minor  structures are  seen in  the
    center  of the  sharpest objects,  which is  acceptable  given the
    quality of the NICMOS PSF.}
  \label{deconv_HST}
\end{figure*}

We determine  the relative  positions of the  lens components  and the
light profile for  the main lens galaxy $G1$  and its closest neighbor
$G2$  (see Fig.~\ref{deconvim})  by analyzing  our HST  images  of the
system.   These data  were obtained  in the  framework of  the CASTLES
program (Cfa-Arizona Space Telescope  LEns Survey), which provides HST
images for  all known  gravitationally lensed quasars.   We deconvolve
these  images  using  a  modified  version of  the  MCS  deconvolution
algorithm   for    images   with    poor   knowledge   of    the   PSF
\citep{magain2007}.  We  previously used  this approach to  unveil the
faint Einstein ring and the lensing  galaxy hidden in the glare of the
quasar   images    of   the   so-called    ``cloverleaf''   HE1413+117
\citep{chantry2007}.   We   analyze  the  Near   Infrared  Camera  and
Multi-Object Spectrometer (NICMOS) F160W (H-band) images obtained with
the NIC2 camera.  The data  consist of four dithered MULTIACCUM images
with  a total  exposure  time of  2752~s  and a  mean  pixel scale  of
0.07568\arcsec  \citep{krist2004}.   We  calibrate  the  images  using
CALNICA, the HST image reduction pipeline, subtract constants for each
quadrant of  NIC2 to give  each image a  mean background of  zero, and
create a noise map for each frame.

The images are simultaneously deconvolved  in a manner similar to that
used for  the EULER and  SMARTS data. The  NICMOS frames lack  any PSF
stars, so we  construct the PSF using the  quasar images themselves in
the iterative  method of  \citet{chantry2007}.  We first  estimate the
PSF  of  each frame  using  Tiny  Tim  \citep{krist2004} and  then  we
deconvolve  them   to  have  the   final  Gaussian  PSF.   During  the
deconvolution, each  image is decomposed  into a set of  point sources
and  any  extended  flux.   The  latter is  then  reconvolved  to  the
resolution of the  original data and subtracted from  the four initial
frames,  leading to  images with  far less  contamination  by extended
structures.  Four new PSFs are estimated from these new images, and we
carry  out a  new deconvolution.  The  process is  repeated until  the
residual maps are  close to being consistent with  the estimated noise
\citep[e.g.][]{courbin1998}.   In this  case,  convergence is  reached
after three iterations and the  final reduced $\chi^{2}$ is 3.59.  The
final deconvolved  image shown  in Fig.~\ref{deconv_HST} has  half the
pixel scale  of the initial images and  a Gaussian PSF with  a FWHM of
0.075\arcsec.

As part of the MCS deconvolution  we also fit analytical models to the
main lens galaxy ($G1$) and  its nearby companion $G2$.  The main lens
is an  early-type galaxy  \citep{eigenbrod2006b}, as its  companion is
likely to be,  so we use elliptical de  Vaucouleurs profiles for both.
The  uncertainties are estimated  by the  scatter of  the measurements
from  a separate  set  of fits  to  each independent  image.  We  also
estimate that there  are systematic errors in the  astrometry from the
NICMOS   pixel   scale   and   focal  plane   distortions   of   order
2~milli-arcseconds based  on our  earlier fits to  the NICMOS  data of
H1413+117 \citep{chantry2007}. These  systematic errors are compatible
with the \citet{lehar2000} comparison of NICMOS  and VLBI astrometry
for radio lenses.

The relative astrometry and photometry of the lens components and of the lensing  galaxies $G1$ and $G2$ are given in Table~\ref{astr_HST}. Coordinates are relative to image $B$, in the same orientation as Fig~\ref{deconv_HST}. The photometry is in the Vega system. For each measurement, we give the $1 \sigma$ internal error bars, to which a systematic error of 2 milli-arcsec should be added.
The models for $G1$ and $G2$ are presented in Table~\ref{gal_HST}, with the effective semi-major and semi-minor axes of the light  distribution $a_{0}$ and $b_{0}$. Each measurement is accompanied by its $1 \sigma$ error bar.

\begin{table}[h!]
  \centering 
  \caption{Relative astrometry and photometry for the four components
    of \obj\  and for the lensing galaxies $G1$ and $G2$.}
  \begin{tabular}{c|ccc}
    \hline
    \hline
    ID & $\Delta \alpha$ (\arcsec) & $\Delta \delta$ (\arcsec) & Magnitude \\ 
    \hline
    B  &  0.                    &  0.                   & 17.77 $\pm$ 0.02 \\
    A1 & $-2.1946$ $\pm$ 0.0004 &  1.2601  $\pm$ 0.0003 & 17.16 $\pm$ 0.02 \\
    A2 & $-1.4809$ $\pm$ 0.0004 &  1.3756  $\pm$ 0.0005 & 17.52 $\pm$ 0.02 \\
    C  & $-2.1128$ $\pm$ 0.0003 &$-0.2778$ $\pm$ 0.0003 & 17.88 $\pm$ 0.02 \\
    G1 & $-1.4388$ $\pm$ 0.0019 &  0.3113  $\pm$ 0.0008 & 18.59 $\pm$ 0.03 \\
    G2 & $-5.4100$ $\pm$ 0.0006 &  0.2850  $\pm$ 0.0003 & 18.14 $\pm$ 0.02 \\
    \hline
  \end{tabular}
  \label{astr_HST}
\end{table}

\begin{table}[h!]
  \centering 
  \caption{Shape  parameters  for   the  main  and  secondary  lensing galaxies. }
  \begin{tabular}{c|cccc}
    \hline
    \hline
    Obj.  & PA   ($^{\circ}$)  & Ellipticity & $a_{0}$ (\arcsec) & $b_{0}$   (\arcsec) \\
    \hline
    G1 & 27.8 $\pm$ 4.3 & 0.18 $\pm$ 0.03 & 0.665 $\pm$ 0.036 & 0.556 $\pm$ 0.025\\
    G2 &  6.4 $\pm$ 3.1 & 0.15 $\pm$ 0.02 & 0.389 $\pm$ 0.004 & 0.334 $\pm$ 0.005\\
    \hline
  \end{tabular}
  \label{gal_HST}
\end{table}

\section{Time delay measurement}
\label{timedelays}

We  measure  the  time  delays  between the  blended  light  curve  of
$A_1$/$A_2$ and  images $B$ and $C$ using  three different techniques:
$i$  the  minimum  dispersion  method of  \citet{pelt1996};  $ii$  the
polynomial  method of  \citet{kochanek2006}; and  $iii$ the  method of
\citet{burud2001}.   Since \obj\  shows  well-defined variations  (see
Fig.~\ref{lightcurves}), it is already clear by visual inspection that
$\Delta t_{B-A} \sim 35$~days and $\Delta t_{B-C} \sim 65$~days.

\subsection{Minimum dispersion method}
\label{Pelt}

\begin{figure}[t]
  \begin{center}
    \leavevmode
    \includegraphics[width=8.5cm]{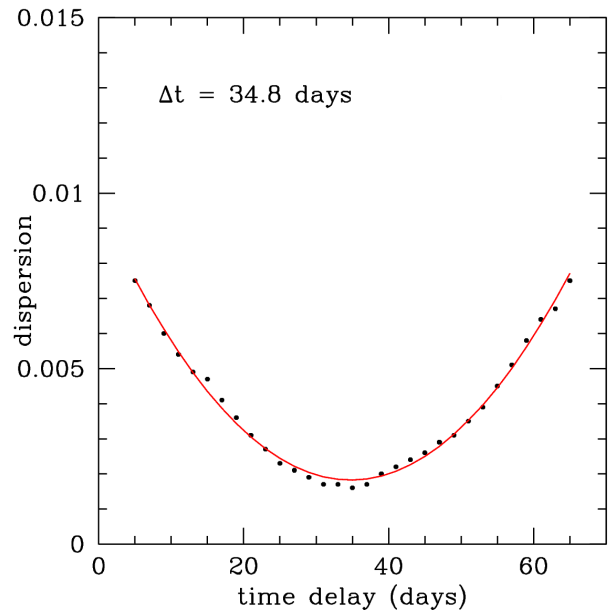}
    \caption{Example  of  a  dispersion  curve as  obtained  from  the
      minimum  dispersion method,  for  components $B$  and $A$.   The
      position  of the parabola minimum gives the time
      delay. Each point  is separated by 2 days,  i.e. about half the data
      mean sampling.  The time delay indicated here is for
      only one realization of the boot-strap procedure (see text).}
    \label{Pelt_example}
  \end{center}
\end{figure}

\begin{figure}[t!]
  \begin{center}
    \leavevmode
    \includegraphics[width=8.5cm]{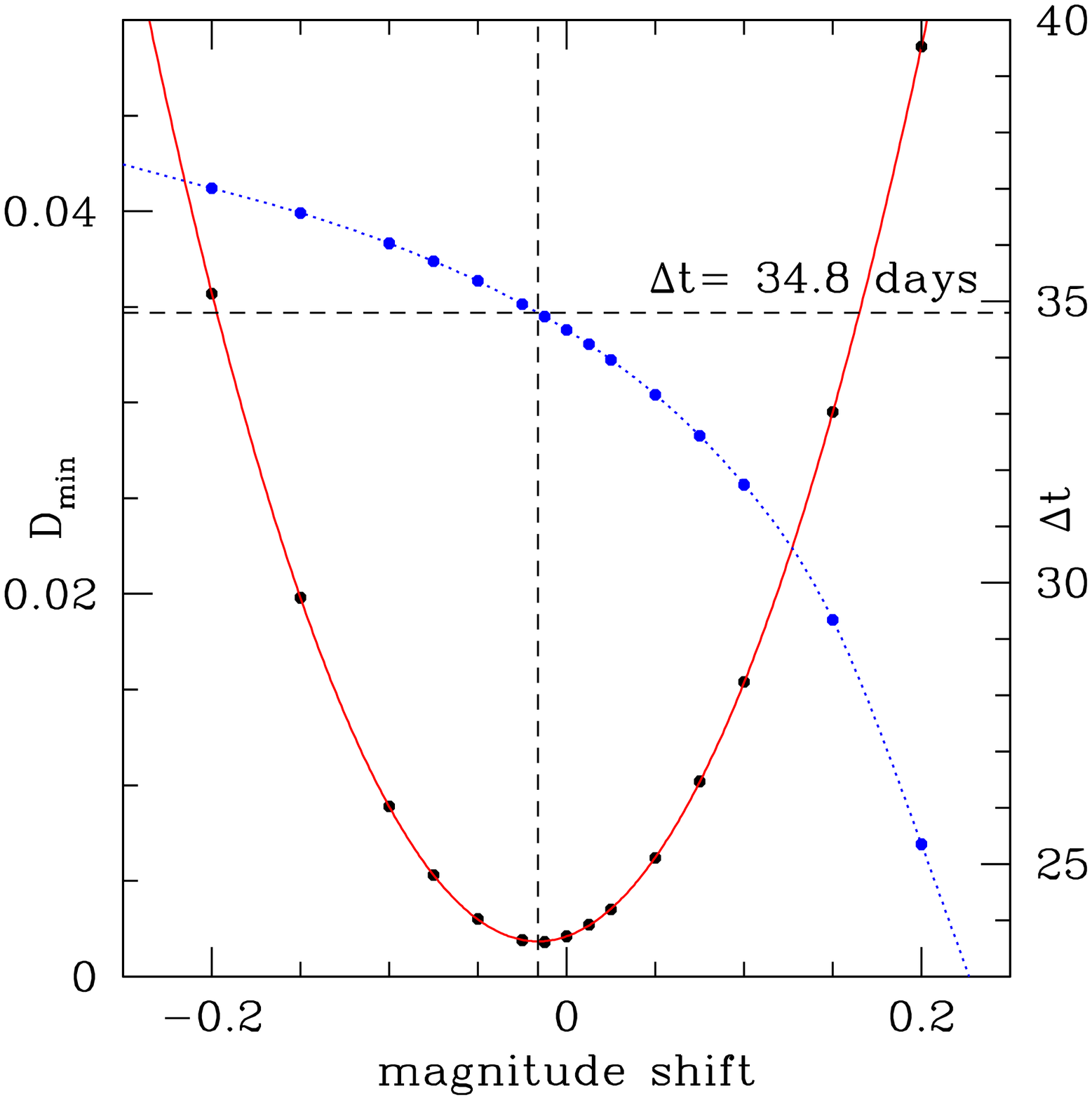}
    \includegraphics[width=8.5cm]{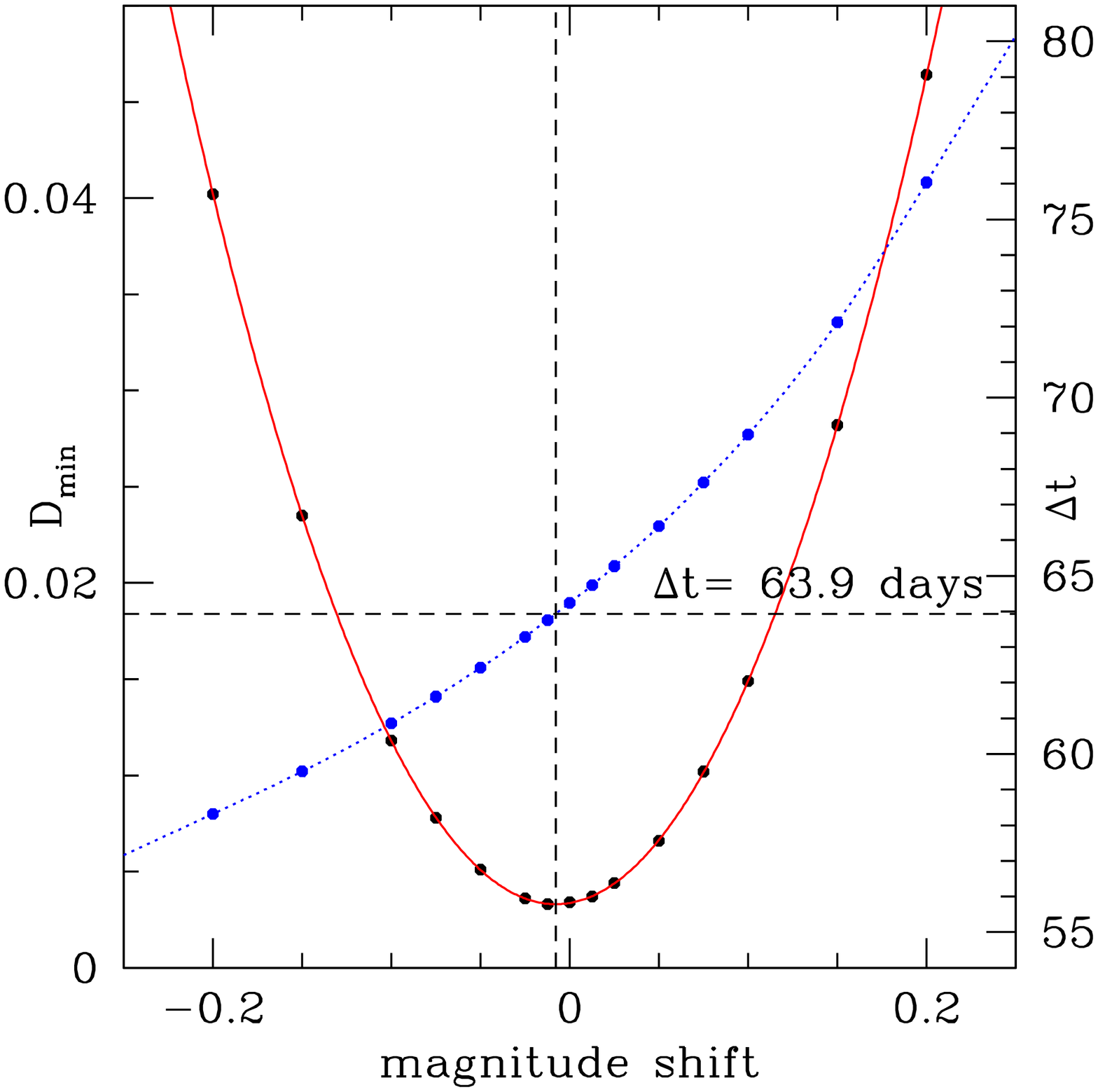}
    \caption{Variation    of   the    dispersion    function   minimum
      $D_{\rm{min}}$  (red  solid  curves),   as  a  function  of  the
      magnitude    shift    used    for   the    normalization    (see
      Sect.~\ref{Pelt}).    Each  $D_{\rm{min}}$   corresponds   to  a
      different  estimate of  the time  delay, indicated  on  the blue
      dotted curves. The  final time delay is the  one with the lowest
      $D_{\rm{min}}$. The top panel is for the $B$-$A$ time delay, the
      bottom one for $B$-$C$.  Time delays indicated here are for only
      one realization of the boot-strap procedure (see text).}
    \label{PeltBABC}
  \end{center}
\end{figure}

In the minimum  dispersion method, time delays are  computed for pairs
of light  curves using a  cross-correlation technique that  takes into
account irregular sampling.  The two light curves are first normalized
to  have zero  mean.   Then, one  of the  light  curves is  used as  a
reference and the second curve is shifted relative to it by a range of
time delays.  For each delay, we calculate the mean dispersion between
the  shifted  points  and  their  nearest temporal  neighbors  in  the
reference light curve.   The best time-delay estimate is  the one that
minimizes this  dispersion function.  Since  the mean sampling  of our
curves is one epoch every four days  and since there is a limit to the
number of time  delays that can be tested  independently, we test time
delays in  steps of 2 days.  Fig.~\ref{Pelt_example}  shows an example
of a  dispersion curve where we have  then fit a parabola  and set the
best time  delay to  be the  one corresponding to  the minimum  of the
parabola.

There is, however, a complication in the step of normalizing the light
curves, arising  from sampling  the light curve  of each  lensed image
over  a different  time  period of  the  intrinsic source  variability
\citep{vuissoz2007}.    We  solve  this   problem  by   computing  the
dispersions as a function of a small magnitude shift $\Delta m$ in the
normalization, measuring  both the minimum  dispersion $D_{min}(\Delta
m)$ and the  best fitting time delay $\Delta t(\Delta  m)$ as shown in
Fig.~\ref{PeltBABC}.   Our  final  value  for  the delay  is  the  one
corresponding  to the  shift  $\Delta m$  that  minimizes the  overall
dispersion.

We  then estimate the  uncertainties by  randomly perturbing  the data
points,  based  on a  Gaussian  distribution  with  the width  of  the
measurement  errors, and  computing  the dispersions  and time  delays
again.  We define the uncertainties by the $1\sigma$ dispersion in the
results  for   100,000  trials  \citep{vuissoz2007}.    The  resulting
uncertainty  estimates are  symmetric  about the  mean,  so our  final
estimates based on this method are
\begin{eqnarray}
 \Delta t_{B-A} =& 35.6 \pm 1.3 ~\ \hbox{days} ~\ (3.6\%) \nonumber \\ 
 \Delta t_{B-C} =& 64.6 \pm 3.4 ~\ \hbox{days} ~\ (5.3\%) 
\end{eqnarray}

While we  have not  taken microlensing effects  into account  for this
analysis, it should matter little, as the method is not very sensitive
\citep{eigenbrod2005}   to  the   very   low  amplitude   microlensing
variability observed for this system (see Sect.~\ref{poly}).

\subsection{Polynomial fit of the light curves}
\label{poly}

\begin{figure}[t]
  \begin{center}
    \leavevmode
    \includegraphics[width=8.5cm]{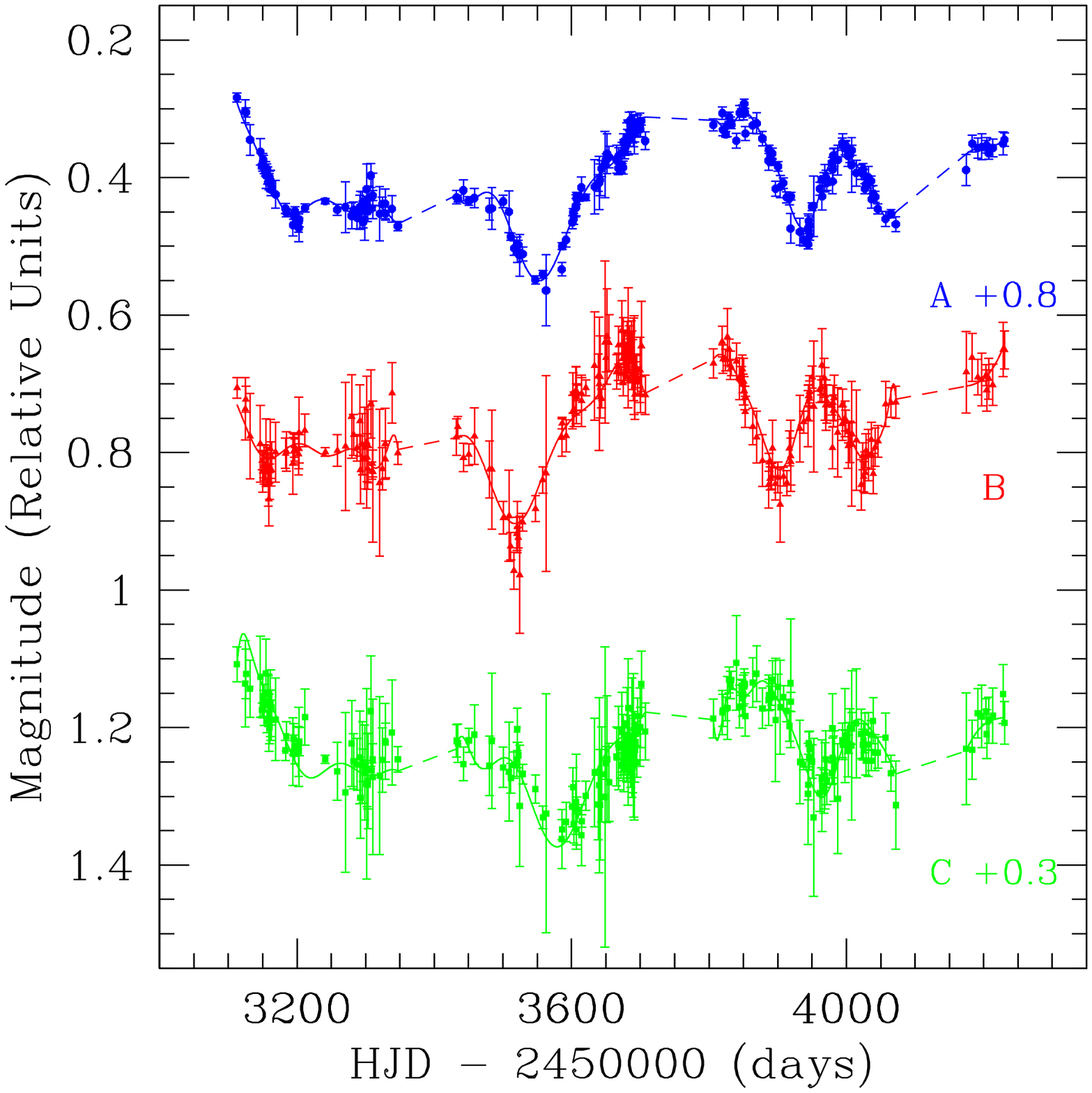}
    \includegraphics[width=8.5cm]{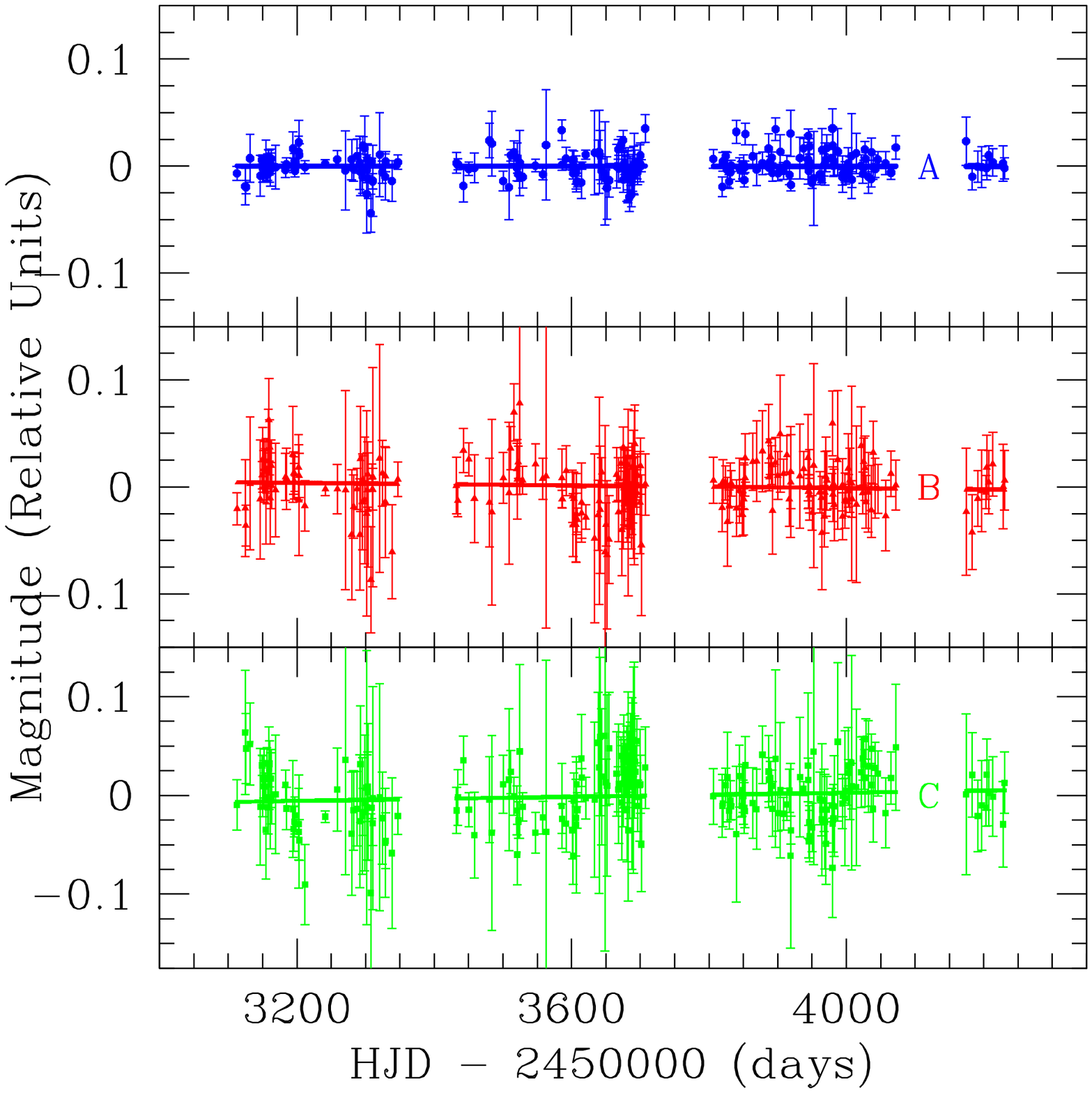}
    \caption{{\it Top:} Best polynomial  fit to the light curves,  
      which are shifted  vertically for display purpose.  
      {\it Bottom:} The residuals of the fit.}
    \label{poly1}
  \end{center}
\end{figure}

In  the polynomial  method \citep{kochanek2006},  the  intrinsic light
curve of the source is  modeled as a polynomial that is simultaneously
fit to all three light curves. Each quasar image has an additional low
order  polynomial  to  model  slow, uncorrelated  variability  due  to
microlensing.  We increase the source polynomial order for each season
until the improvement in the  $\chi^2$ statistics of the fits cease to
be significant.  This results in using polynomial orders of 11, 10, 17
and 4  for the  four seasons of  data. The low  amplitude microlensing
variations  are modeled  with a  simple linear  function for  the four
seasons.  Fig.~\ref{poly1}  shows the  best fits to  the data  and the
residuals from the model.  The  effects of microlensing in this system
are very  small, with  variations of only  $\sim 0.01$~mag  over three
years.   As  with  the  minimum  dispersion method,  we  estimate  the
uncertainties by  randomly perturbing  the light curves  100,000 times
and using the standard deviation  of the trials as the error estimates
to find that
\begin{eqnarray}
 \Delta t_{B-A} =& 35.0 \pm 1.1 ~\ \hbox{days} ~\ (3.0\%)  \nonumber \\
 \Delta t_{B-C} =& 61.2 \pm 1.5 ~\ \hbox{days} ~\ (2.4\%)
\end{eqnarray}

\subsection{Numerical modeling of the light curves}
\label{num}

\begin{figure}[t]
  \begin{center}
    \leavevmode
    \includegraphics[width=8.5cm]{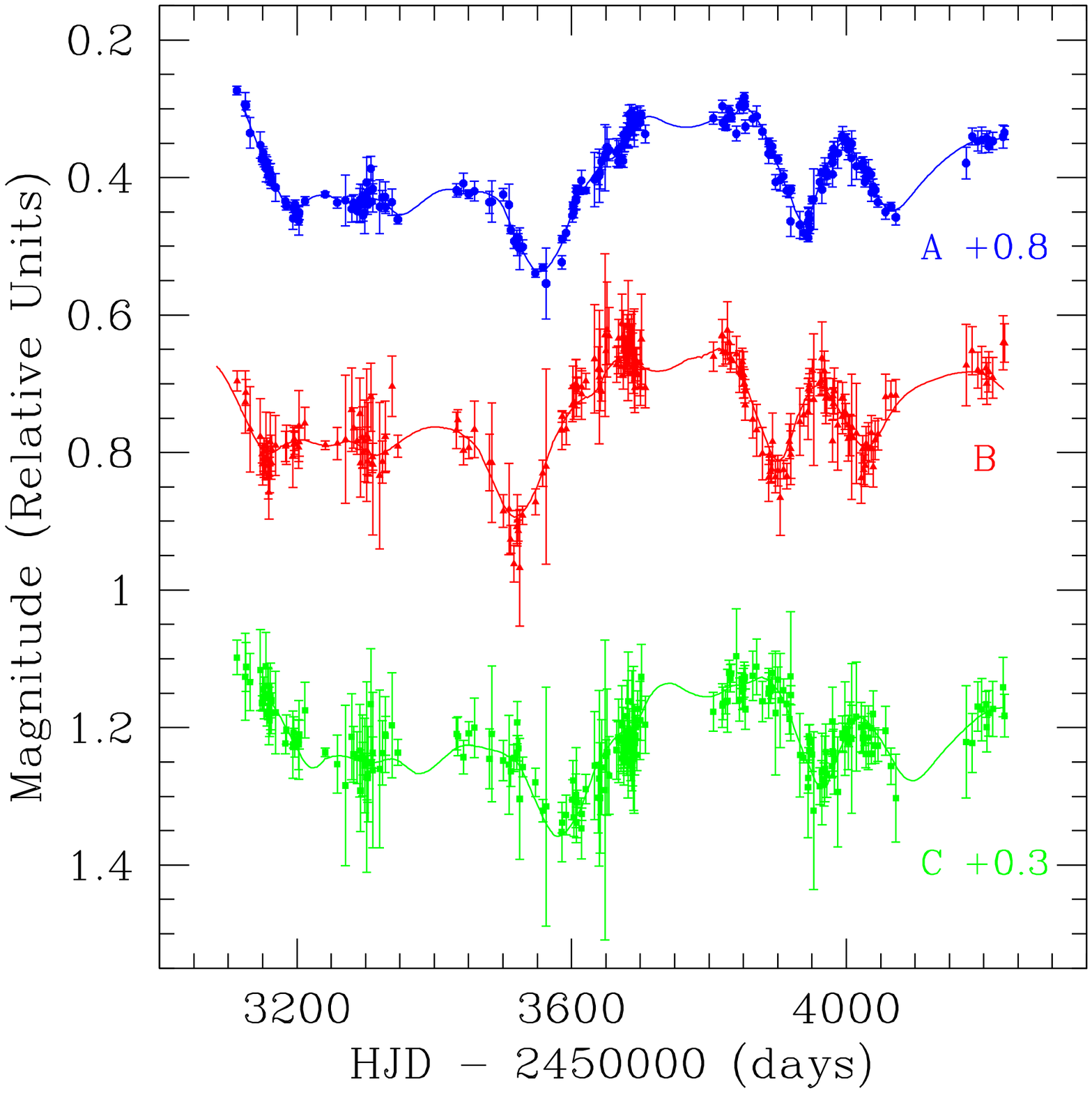}
    \includegraphics[width=8.5cm]{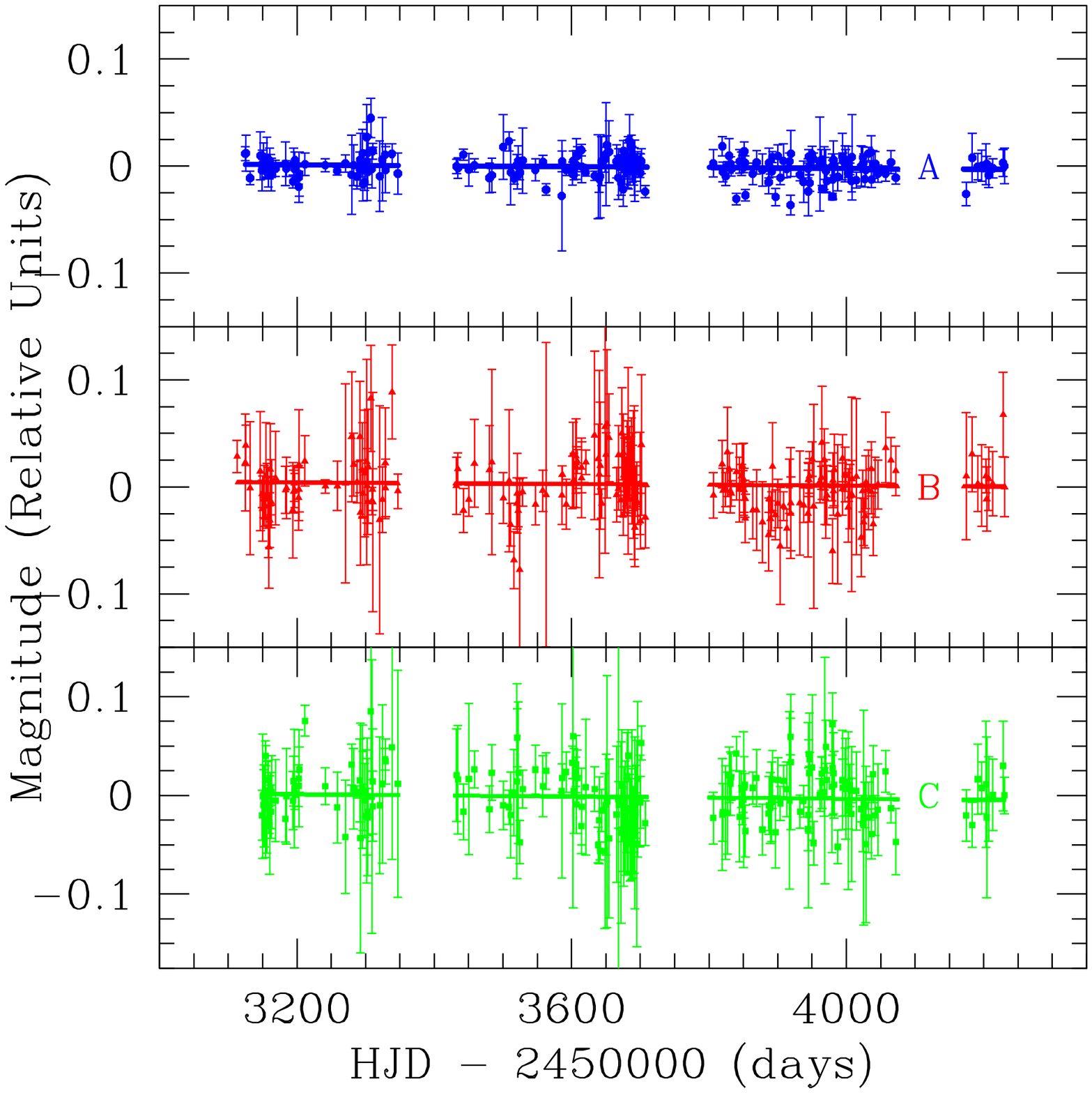}
    \caption{{\it Top}:  The light curves  of the quasar  images shown
      along with the best numerical model.  
      {\it Bottom}: The residuals of the fit.}
    \label{numerical}
  \end{center}
\end{figure}

Our   last   approach   is   based   on  the   method   described   in
\citet{burud2001}, which determines the time delay between a
pair of light  curves using a gridded numerical  model for the source
light curve.  For a series of  time delays, we fit the data with
a  flux  ratio  between  the  two  curves,  and  a  linear  trend  for
microlensing on each full light  curve.  The numerical source model is
smoothed on the  scale of the temporal sampling,  based on a smoothing
function weighted  by a Lagrange  multiplier.  The best time  delay is
the  one that  minimizes the  $\chi^{2}$  between the  model and  data
points.

This method has several advantages: first, none  of the data
light curves is taken as a reference: they are all treated on an equal
basis.  Furthermore,  as the model is purely  numerical, no assumption
is made  on the  shape of the  quasar's intrinsic light  curve (except
that  it is  sufficiently smooth).  Finally,  a model  light curve  is
obtained  for the  intrinsic  variations  of the  quasar,  as for  the
polynomial fit method (see Sect.~\ref{poly}).

We  have applied  this method  to  the two  pairs of  light curves  of
\obj. The resulting  fits to the light curves  and their residuals are
shown in Fig.~\ref{numerical}.  Using a Monte Carlo method to estimate
the  uncertainties,   we  find  from  7,000   trials  (adding  normally
distributed random  errors with the appropriate  standard deviation on
each data point) :
\begin{eqnarray}
 \Delta t_{B-A} =& 36.0  \pm 1.5  ~\ \hbox{days} ~\ (4.2\%) \nonumber \\
 \Delta t_{B-C} =& 61.9~ ^{+\,6.7}_{-\,0.5}  ~\ \hbox{days} ~\ (^{+\,11\%}_{-\,1\%}) 
\end{eqnarray}

We note a  secondary peak  in the  $\Delta t_{B-C}$  Monte Carlo
distribution,  around  69~days, in  addition  to  the  main peak  at
61.9~days.  There is, however, no  evidence of such a secondary peak
in the results  of the minimum dispersion method  and the polynomial
fitting technique.   This translates into an  asymmetrical error bar
on  the result  obtained with  the  numerical fitting  of the  light
curves, and is taken into account  in our final estimate of the time
delay between quasar images $B$ and $C$.

\subsection{Final time delays}

\begin{figure}[t!]
  \begin{center}
    \includegraphics[width=8.5cm]{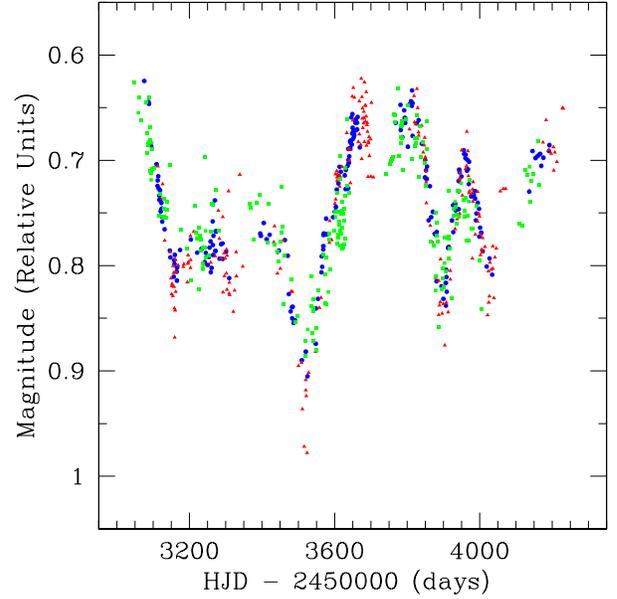}
    \caption{Light curves of the three quasar images, shifted by their
      respective  time  delay  and   flux  ratio.   The  blue  circles
      correspond to image $A$, the  red triangles to $B$ and the green
      squares to $C$.}
    \label{poly3}
  \end{center}
\end{figure}

For the final delay estimate we adopt the unweighted mean of the three
methods,  and we  take  as  uncertainties the  quadrature  sum of  the
average statistical error  and the dispersion of the  results from the
individual methods about  their mean.  Our final estimate  of the time
delays is
\begin{eqnarray}
 \Delta t_{B-A} =& 35.5  \pm 1.4 ~\ \hbox{days} ~\ (3.8\%) \nonumber \\
 \Delta t_{B-C} =& 62.6~ ^{+\,4.1}_{-\,2.3} ~\ \hbox{days} ~\ (^{+\,6.5\%}_{-\,3.7\%})
\end{eqnarray}

We  cannot measure  the time  delay between  the individual  $A_1$ and
$A_2$ light  curves, but values larger  than $\Delta t_{A1-A2}=2$~days
are incompatible with  any of the models we  consider in the following
section.  We  can nevertheless estimate  the flux ratio  between $A_1$
and $A_2$.   After correcting  for the time  delays, we find  that the
$R$-band flux ratios between the images are
\begin{equation}
  \frac{F_{A}}{F_{B}} = 2.88 \pm 0.04, ~\ 
  \frac{F_{A}}{F_{C}} = 3.38 \pm 0.06, ~\ 
  \frac{F_{A1}}{F_{A2}}= 1.37 \pm 0.05
\end{equation}

Fig.~\ref{poly3}  displays the  light  curves obtained  for the  three
quasar  images, after  shifting by  the time  delays and  flux ratios.
Note  that these  flux  ratios  differ  from those  measured  by
\citet{morgan2004} from the MgII  broad emission lines, probably due
to  long-term microlensing (on  a longer  scale than  our monitoring
3-year baseline), as discussed in the next section.

\begin{figure}[t!]
  \begin{center}
    \leavevmode
    \includegraphics[width=8.9cm]{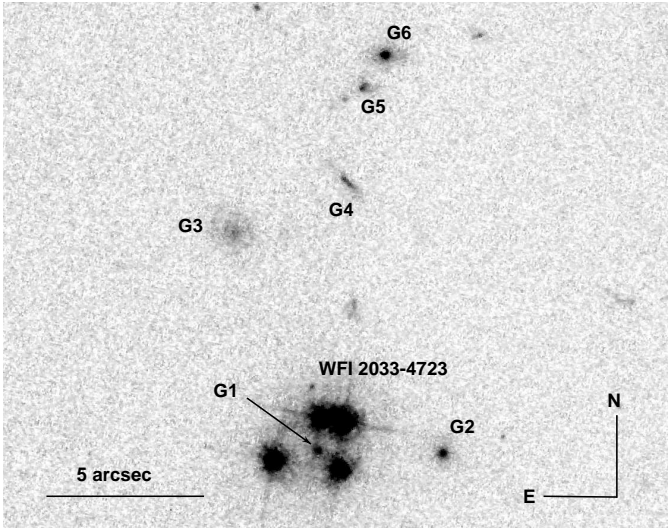}
    \caption{Environment of \obj\ as  seen in an HST/ACS F814 (I-band)
      image.  The main  lens galaxy $G1$ and the  close companion $G2$
      were included in  our analysis of the NICMOS  image, and here we
      have labeled additional group members as $G3$-$G6$.}
    \label{WFI_HST}
  \end{center}
\end{figure}

\begin{table*}[ht!]
  \centering   \caption{Result  of   the  parametric   lens  modeling.} 
  \begin{tabular}{lc|ccccccl} \hline \hline Name  & Comp.  & mass & e,
  $\theta_e$ & $\gamma$, $\theta_{\gamma}$ & \#d.o.f. & $\chi^2$ & $h$
  & Comments\\
\hline
 SIE+$\gamma$ &  & b=0.96 & 0.21, (20.8) & 0.187, 7.4  & 1 & 15.3 &  -   &  Time delays not used\\ 
 SIE+$\gamma$ &  & b=0.94 & 0.13, (30.5) & 0.063, 24.6 & 2 & 3.6  & $0.79^{+0.04}_{-0.02}$ & $G_{\rm group}$ included \\ 
 SIE+$\gamma$ &  & b=0.97 & 0.16, 84.4   & 0.059, 46.9 & 1 & 0.30 & $0.63^{+0.07}_{-0.03}$ & $G_{\rm group}$ included  \\ 
 dVC+$\gamma$ &  & b=2.71 & 0.20, (20.1) & 0.305, 9.7  & 1 & 34.4 &  -   &  Time delays not used\\
 dVC+$\gamma$ &  & b=2.83 & 0.18,  83.1  & 0.116, 64.5 & 1 & 0.01 & $0.92^{+0.06}_{-0.03}$ & $G_{\rm group}$ included    \\
 NFW+$\gamma$ &  & $\kappa_s$=0.20 & 0.16, (27.4) & 0.070, -3.6 & 1 & 0.38 & - & $r_s$=10\arcsec (fixed); time delays not used\\
 NFW+$\gamma$ &  & $\kappa_s$=0.21 & 0.15, 85.7 & 0.079, 9.5 & 1 & 0.06 & $0.29^{+0.03}_{-0.03}$ & $r_s$=10\arcsec (fixed); $G_{\rm group}$ included\\
 NFW+$\gamma$ &  & $\kappa_s$=0.09 & 0.15, 85.4 & 0.076, 30.6 & 1 & 0.01 & $0.63^{+0.10}_{-0.08}$ & $r_s$=1\arcsec (fixed); $G_{\rm group}$ included\\
 dVC+NFW+$\gamma$& Light & b=1.56 & (0.17), (29.3) & 0.057, 37.0 & - & -  & -  & $R_e=0.608$\arcsec (fixed)\\ 
                 & Halo  & $\kappa_s$=0.082 & 0.065, (29.3) &   & 1 & 6.33 & $0.78^{+0.12}_{-0.10}$ & $r_s$=10\arcsec (fixed); $G_{\rm group}$ included\\ 
dVC+NFW+$\gamma$ & Light & b=1.53 & (0.16), (26.4) & 0.075, 27.5 & - & -  & -  & same model as above, with\\ 
                 & Halo & $\kappa_s$=0.10 & 0.43, 89.8 &   & 3 & 3.2 & 0.69$^{+0.20}_{-0.10}$ & flux ratios included \\ 
\hline
  \end{tabular}
  \label{mod}
\end{table*}

\section{Parametric modeling}
\label{param_models}

\subsection{Observational constraints}

We constrain the mass models of  \obj\ using the positions of the four
lensed images, the position of the  lens galaxy $G1$ and the two delay
measurements,  for  a  total  of  12  constrains.   Except  when
indicated, we do not use the  image flux ratios because they can be
affected by  extinction \citep{falco1999, jean1998}  and milli-lensing
by substructures \citep{mao1998}.  We can also constrain the structure
of  $G1$ given  its  ellipticity $e$,  position  angle $\theta_e$  and
effective  radius  $R_e$  to  the  extent that  these  properties  are
correlated with  those of  its dark matter  halo. Although  a possible
mismatch between  the light and mass distributions  is not impossible,
we adopt  the formal  errors of 0.002\arcsec\  on the position  of the
lens $G1$ (Table~\ref{gal_HST}). This is motivated by the small offset
between the  centers of mass and  light found by  \citet{yoo2006} in a
sample of four lensing galaxies.

Finally, \obj\ is located in a group of galaxies, labeled $G2$-$G6$ in
Fig.~\ref{WFI_HST}.  We  include $G2$ as a  singular isothermal sphere
(SIS) in  all our models since  it is close (4\arcsec)  and of similar
luminosity  to   $G1$.  As   \citet{morgan2004},  we  are   unable  to
successfully  model the  system without  including $G2$.   When enough
observational  constraints  are  available  we  further  add  galaxies
$G3$-$G6$ as a SIS mass distribution located at the barycenter $G_{\rm
  group}$ of the group.  In all models we include an external shear of
amplitude $\gamma$ and  position angle $\theta_\gamma$ that represents
the   gravitational  perturbations   due  to   mass   unaccounted  for
explicitly.  We also experiment with including mass at the position of
object $X$ (Fig.~\ref{deconv_HST}, 2\arcsec\  North of $G1$) and find
that doing so does not improve the models.

We consider a  sequence of standard mass models  for $G1$, including a
singular isothermal ellipsoid (SIE),  a de Vaucouleurs (dVC) model and
a  NFW model \citep{NFW}, and we fit  the data  using \texttt{LENSMODEL}
(v1.99g)  \citep{keeton2001}. 
The results are summarized  in Table~\ref{mod}, where columns $\#1$ and $\#2$ describe the model family,  and $\#3$ the mass parameter \citep[either the Einstein  radius $b$  in arcseconds or  the mean  mass surface density $\kappa_s$, as defined in][]{keeton2001}. Column $\#4$ is for the ellipticity $e$ and PA $\theta_e$ of the lens  $G1$. Note that the measured PA of $G1$ is $\theta = 27.8^{\circ}$ (Table~\ref{gal_HST}).  Column $\#5$ gives the external shear amplitude $\gamma$ and PA $\theta_{\gamma}$, $\#6$ the number of degrees of freedom for each model, and $\#7$ the resulting reduced $\chi^2$. Column $\#8$ finally shows the best estimate for $h=$\ho$/100$.  A minus sign in this column means that time delays are not used as constraints.  All angles are given positive East of North, and values given in parentheses are fitted to the observations. All models assume  $\Delta   t_{A1-A2}  =  2$~days and include the companion galaxy $G2$, with a resulting mass $0.1\, m_{G1} < m_{G2} < m_{G1}$.   

\subsection{SIE Models}

Our first model  consists of an SIE  for $G1$, an SIS for  $G2$ and an
external shear.   When we fit only  the image positions  but include a
prior on the position  angle $\theta_e$ (from Table~\ref{astr_HST}) we
do not  find a good fit unless  the constraint on the  position of the
lensing  galaxy  is  relaxed.  The  prior on  the  position  angle  is
justified  by  statistical studies  finding  correlations between  the
position angles but not the axis  ratios of the visible and total mass
distributions \citep{ferreras2008, keeton1997}.  With the inclusion of
the time  delays we have enough  constraints to add the  group halo to
the model.   With the  position angle of  $G1$ constrained,  we obtain
poor fits to  the data with reduced $\chi^2=3.6$  for $N_{\rm dof}=2$.
When we leave the position angle free, we find good fits but the model
PA  is  $55^\circ$  from   the  observed.  These  models  have  Hubble
parameters of $h\simeq 0.63^{+0.07}_{-0.03}$ with the spread dominated
by the degeneracies between the ellipticity and the shear.

\subsection{De Vaucouleurs Models}

Next  we consider  a constant  mass-to-light ratio  model of  the lens
galaxy based  on a  De Vaucouleurs model.  The position angle  and the
effective  radius  $R_e=0.608$\arcsec\  (the  geometric  mean  of  the
semi-axes   in  Table~\ref{astr_HST},   corresponding  to   4~kpc  for
$h=0.72$) are constrained by the values measured for the galaxy.  This
model does  not fit well  the lens configuration  ($\chi^2\sim 34.4$),
mainly due to the small  uncertainty on the lens galaxy position. When
we include the time delays we find a good fit ($\chi^2\simeq 0.01$) as
long as  we allow $G1$ to  be misaligned with respect  to the observed
galaxy.  As expected from the  reduced surface density compared to the
SIE model  \citep{kochanek2002}, we find  a much higher value  for the
Hubble parameter, $h=0.92$.

\subsection{NFW Models}

We   use   an    NFW   model   with   a   fixed    break   radius   of
$r_s=10$\arcsec\ (40~kpc),  where the break  radius is related  to the
virial  radius through the  concentration $c=R_{\rm  vir}/r_s$.  Since
$r_s$  lies  well  outside  the  Einstein  radius  of  the  lens,  its
particular value  is not  important.  This model  is not  realistic by
itself because the shallow $\rho  \propto 1/r$ central density cusp of
the model will lead to a visible central image.  We again find that we
can fit the  astrometry well even when the position  angle of the lens
is constrained,  but we cannot do  so after including  the time delays
unless we  allow the model  of $G1$ to  be misaligned relative  to the
light. In  any case,  this model leads  to a  fifth image about  3 mag
fainter than $A$  that should be visible on our  NICMOS data. This NFW
model has a higher surface density near the Einstein ring than the SIE
model, so we find a lower  value for the Hubble parameter of $h \simeq
0.29$.     Using    an   unphysically    small    break   radius    of
$r_s=1$\arcsec\ raises  the density and hence the  Hubble parameter to
$h\simeq 0.63$.

\subsection{De Vaucouleurs plus NFW Models}

As our final, and most  realistic, parametrized model we combine a dVC
model  constrained by the  visible galaxy  with an  NFW model  for the
halo.  The two components are first constrained to have the same
position  and position  angle, the  parameters  of the  dVC model  are
constrained  by the  observations,  and  the NFW  model  has a  fixed
$r_s=10$\arcsec\ break radius.   This model leads to a  poor fit, with
$\chi^2=6.33$ for  $N_{\rm dof}=1$.   When we free  the PA of  the NFW
model,  we  find  an  acceptable  fit for  $N_{\rm  dof}=0$,  but  the
misalignment of the NFW model relative to the optical is $40^{\circ}$.

We  can further constrain the  model by including  the three MgII
emission line flux ratios  from \citet{morgan2004}.  We use the line
flux ratios instead of those obtained from the light curves, because
they  should  be  insensitive  to microlensing.   With  these  three
additional  constraints  we   still  find  that  the  $89.8^{\circ}$
position  angle of  the NFW  model is  strongly misaligned  from the
$26.4^{\circ}$  position  angle  of  the  dVc  model,  indicating  a
twisting of the mass isocontours.   The model has a reduced $\chi^2$
of  $3.2$ for  $N_{\rm dof}=3$.   The model  flux  ratios are
significantly different from  the constraints.  We find $F_{A}/F_{B}
=  2.81$, $F_{A}/F_{C}  =  5.01$, and  $F_{A1}F_{A2}  = 1.26$  while
\citet{morgan2004} report $F_{A}/F_{B}  = 2.55 \pm 0.60$, $F_{A}/F_{C}
= 2.02 \pm 0.35$, and $F_{A1}/F_{A2}  = 1.61 \pm 0.35$.  The match of
the flux ratios is better if  we do not include the constraints from
the  time   delays.   In  all   cases,  $F_{B}/F_{C}$  is   the  most
``anomalous''     flux      ratio,     as     also      found     by
\citet{morgan2004}.  
While  the differences between  the line  and continuum  flux ratios
suggests the presence of  long-term microlensing, we see no evidence
for  the   time  variability  in  the  flux   ratios  expected  from
microlensing.  We  also note  that the broad  line flux  ratios vary
with  wavelength   \citep{morgan2004},  which  suggests   that  dust
extinction may as well be affecting the flux ratios.

\begin{figure}[t!]
  \begin{center}
    \leavevmode
    \centering
    \includegraphics[width=6.5cm]{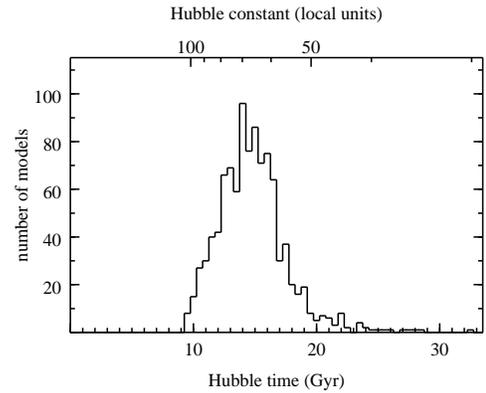}
    \caption{Distribution of \ho\ from 1000 non-parametric models. The
      bottom-axis gives the Hubble time, the
      top-axis \ho\ in \kmsmpc. }
    \label{Ho_saha}
  \end{center}
\end{figure}

\section{Non-parametric modeling}
\label{saha_models}

\begin{figure}[t!]
  \begin{center}
    \leavevmode
    \centering
    \includegraphics[width=5.cm]{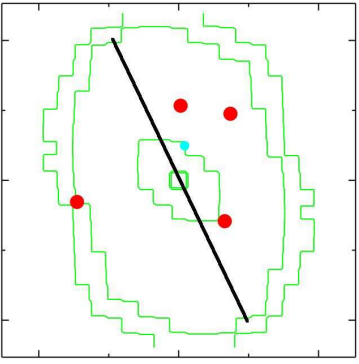}
    \caption{Mean mass distribution  from 1000 pixellated lens models.
      The red dots  are the quasar images and the  blue dot the source
      position.   The thick  solid line  indicates the  observed major
      axis of  the lensing galaxy.  Each  tick-mark measures 1\arcsec.
      The third contour from the outside traces the critical mass
      density        $\Sigma_{\rm         crit}        =        1.19
      \,10^{11}$~M$_{\odot}$\,arcsec$^{-2}$,  and   the  others  are
      drawn logarithmically  from the critical one  (each contour is
      2.5 larger/smaller  than the previous one).  North  is to the
      top and East to the left.}
    \label{mass_saha}
  \end{center}
\end{figure}

We use the non-parametric \texttt{PixeLens} \citep{saha2004} method as
our  second probe  of the  mass distribution.   This approach  has the
advantage that it makes fewer  assumptions about the shape of the $G1$
than the ellipsoidal parametric  models.  The models include priors on
the steepness of the  radial mass profile, imposes smoothness criteria
on  the profile and  we restrict  to models  symmetric about  the lens
center.  We include two point  masses to represent $G2$ and the group.
We run 1000 trial models  at a resolution of $\sim 0.23 \arcsec$/pixel
which are constrained  to fit the image positions  and the time delays
exactly.  For  each model we vary  the Einstein radii of  $G2$ and the
group  over  the  ranges  0.03\arcsec\  $<  R_E(G2)  <$  3\arcsec  and
0.3\arcsec\ $< R_E({\rm group}) <$ 5\arcsec, respectively.  Apart
from the inclusion of these  additional point masses, the method and
priors  are as  explained in  detail in  \citet{coles2008}.  A test,
where the  technique is used to  infer \ho\ from a  N-body and hydro
simulated lens, and an additional discussion of the priors are described in
\citet{read2007}. Fig.~\ref{Ho_saha}   shows    the   resulting
probability distribution  for \ho\ from  the 1000 models.   The median
value of the distribution is
\begin{equation}
\hbox{\ho}  = 67.1~^{+13.0}_{-9.9} ~\ \hbox{\kmsmpc}
\label{coles}
\end{equation}
where the error bars are at 68\% confidence. As already illustrated by
our  parametric modeling,   the predicted  \ho\  value depends  on the
density gradient of the models.
Fig.~\ref{mass_saha} shows  the mean  surface density contours  of the
models, and  we see a twisting of  the contours away from  that of the
visible galaxy in the outer regions.

\section{Conclusions}
\label{conclusion}

By combining data  from COSMOGRAIL and the  SMARTS 1.3m telescope
we measure two independent time  delays in the quadruply imaged quasar
\obj\ \citep{morgan2004}.  The  fractional uncertainties of $\sim 4\%$
are among  the best  obtained so far  from an optical  monitoring.  We
obtain  the  light   curves  of  the  quasar  images   using  the  MCS
deconvolution photometry algorithm \citep{magain1998} and then measure
time delays using three different approaches with a final estimate of
\begin{eqnarray}
 \Delta t_{B-A} =& 35.5  \pm 1.4 ~\ \hbox{days} ~\ (3.8\%) \nonumber \\
 \Delta t_{B-C} =& 62.6~ ^{+\,4.1}_{-\,2.3} ~\ \hbox{days} ~\ (^{+\,6.5\%}_{-\,3.7\%})
\end{eqnarray}
\noindent where $A$  is the mean light curve of  the blended of quasar
images $A_1$  and $A_2$.  We  find little evidence of  microlensing in
this system, which makes \obj\  a very good system for measuring clean
time delays.

The parametric models are   consistent with concordance value  of \ho\
when the lens galaxy has an isothermal mass profile  out to the radius
of  the Einstein ring.  As  expected,  the models allow higher (lower)
values    as the mass   distribution   is more centrally  concentrated
(extended) using  de Vaucouleurs  (NFW) mass distribution.    The
non-parametric models predict \ho\ = 67.1 $^{+13.0}_{-9.9}$ \kmsmpc.

The addition  of the  time delays  as a  constraint on  the lens
models  does  not  alter  the  mismatch  between  the  observed  and
predicted image flux ratios.  The  largest flux ratio anomaly is the
45\% difference between  the MgII flux ratios found  for images B/C.
\citet{morgan2004} also noted that the $F_B$/$F_C$ flux ratio varies
with wavelength, suggesting  the presence of chromatic microlensing.
The lack of significant variability in the flux ratio over our three
year  monitoring period  suggests either  that the  effective source
velocities in this lens are very low or that the affected images lie
in  one of  the broad  demagnified valleys  typical  of microlensing
magnification patterns for low stellar surface densities.

Several galaxies close to the  line of sight have a significant impact
on the mass modeling. We generally model the potential as the sum of a
main   lensing  galaxy   $G1$,   a  companion   galaxy  $G2$   ($\sim$
4\arcsec\ West of $G1$), and a nearby group ($\sim$ 9\arcsec\ North of
$G1$).  Both the parametric and non-parametric models suggest that the
isodensity contours of $G1$ itself must be twisted, with some evidence
that the outer  regions are aligned with the tidal  field of the group
rather than the  luminous galaxy.  This could indicate  that $G1$ is a
satellite   rather   than   the    central   galaxy   of   the   group
\citep[e.g.][]{kuhlen2007}.   The twisting seems  to be  required even
though the angular structure of  the potential can be adjusted through
the companion galaxy $G2$, an  external tidal shear, and in some cases
a group halo.  Clarifying this issue requires more constraints such as
detailed  imaging of  the  Einstein  ring image  of  the quasar  host,
measuring  the redshifts  of the  nearby galaxies,  and  measuring the
velocity dispersion of $G1$.

\begin{acknowledgements}
  We would like to thank all the observers at the EULER telescope for
  the acquisition of these monitoring data.  We also thank Profs.  G.
  Burki and M.  Mayor for their help in the flexible
  scheduling of the observing runs, improving the necessary regular
  temporal sampling.  We are very grateful to Prof. A. Blecha and the
  technical team of the Geneva Observatory for their efficient help
  with the electronics of the CCD camera.  Many thanks also to
  Sandrine Sohy for her commitment in the programming part of the
  deconvolution techniques.  Finally, we would like to thank Chuck
  Keeton for his advice on the use of \texttt{LENSMODEL}.  This work
  is supported by ESA and the Belgian Federal Science Policy Office
  under contract PRODEX 90195.  CSK is funded by National Science
  Foundation grant AST-0708082. COSMOGRAIL is financially supported
  by the Swiss National Science Foundation (SNSF).
\end{acknowledgements}

\bibliographystyle{aa}
\bibliography{bibtex.bib}

\longtab{4}{
\begin{longtable}{ccccccccc}
\caption{\label{data} Photometry of \obj, as in Fig.~\ref{lightcurves}. The Julian date corresponds to HJD-2450000 days.}\\
\hline
\hline
HJD & seeing ["] & mag $A$ & $\sigma_A$ & mag $B$ & $\sigma_B$ & mag $C$ & $\sigma_C$ & telescope \\
\hline 
\endfirsthead 
\caption{continued.}\\ 
\hline
\hline
HJD & seeing ["] & mag $A$ & $\sigma_A$ & mag $B$ & $\sigma_B$ & mag $C$ & $\sigma_C$ & telescope \\
\hline 
\endhead 
\hline
\endfoot
3112.86 & 1.55 & 1.330 & 0.006 & 1.752 & 0.016 & 2.155 & 0.027 & SMARTS \\
3124.85 & 1.58 & 1.350 & 0.017 & 1.783 & 0.045 & 2.183 & 0.063 & SMARTS \\
3125.84 & 1.71 & 1.352 & 0.006 & 1.769 & 0.019 & 2.169 & 0.036 & SMARTS \\
3131.83 & 1.38 & 1.392 & 0.022 & 1.823 & 0.061 & 2.190 & 0.041 & SMARTS \\
3146.90 & 1.45 & 1.409 & 0.018 & 1.834 & 0.053 & 2.173 & 0.056 & SMARTS \\
3149.34 & 1.69 & 1.430 & 0.005 & 1.859 & 0.015 & 2.221 & 0.013 & EULER \\
3150.32 & 1.22 & 1.425 & 0.009 & 1.874 & 0.014 & 2.202 & 0.016 & EULER \\
3151.35 & 1.64 & 1.421 & 0.006 & 1.866 & 0.016 & 2.211 & 0.017 & EULER \\
3152.29 & 1.47 & 1.432 & 0.009 & 1.864 & 0.029 & 2.214 & 0.031 & EULER \\
3154.30 & 1.42 & 1.439 & 0.006 & 1.876 & 0.011 & 2.210 & 0.014 & EULER \\
3154.84 & 1.42 & 1.443 & 0.022 & 1.847 & 0.055 & 2.168 & 0.056 & SMARTS \\
3156.27 & 1.36 & 1.434 & 0.007 & 1.853 & 0.017 & 2.238 & 0.016 & EULER \\
3157.39 & 1.23 & 1.444 & 0.012 & 1.867 & 0.026 & 2.196 & 0.026 & EULER \\
3158.36 & 1.59 & 1.454 & 0.008 & 1.887 & 0.037 & 2.208 & 0.037 & EULER \\
3159.29 & 1.40 & 1.453 & 0.012 & 1.915 & 0.040 & 2.240 & 0.043 & EULER \\
3160.32 & 1.23 & 1.446 & 0.009 & 1.889 & 0.026 & 2.246 & 0.024 & EULER \\
3161.37 & 1.37 & 1.451 & 0.015 & 1.841 & 0.043 & 2.211 & 0.049 & EULER \\
3162.33 & 1.11 & 1.464 & 0.009 & 1.871 & 0.024 & 2.236 & 0.029 & EULER \\
3163.22 & 1.34 & 1.460 & 0.008 & 1.853 & 0.021 & 2.217 & 0.025 & EULER \\
3168.93 & 1.62 & 1.471 & 0.020 & 1.846 & 0.043 & 2.235 & 0.060 & SMARTS \\
3183.37 & 1.78 & 1.491 & 0.007 & 1.848 & 0.022 & 2.280 & 0.014 & EULER \\
3184.78 & 1.28 & 1.498 & 0.006 & 1.845 & 0.029 & 2.260 & 0.021 & SMARTS \\
3194.30 & 1.12 & 1.516 & 0.016 & 1.863 & 0.046 & 2.282 & 0.050 & EULER \\
3195.28 & 1.12 & 1.501 & 0.006 & 1.845 & 0.018 & 2.264 & 0.020 & EULER \\
3196.36 & 1.32 & 1.504 & 0.008 & 1.837 & 0.018 & 2.273 & 0.017 & EULER \\
3197.18 & 1.26 & 1.495 & 0.006 & 1.843 & 0.015 & 2.281 & 0.018 & EULER \\
3202.41 & 1.81 & 1.511 & 0.015 & 1.849 & 0.031 & 2.282 & 0.037 & EULER \\
3202.76 & 1.40 & 1.520 & 0.020 & 1.818 & 0.051 & 2.275 & 0.056 & SMARTS \\
3203.41 & 1.12 & 1.508 & 0.009 & 1.841 & 0.017 & 2.267 & 0.017 & EULER \\
3211.74 & 1.20 & 1.491 & 0.006 & 1.814 & 0.023 & 2.231 & 0.040 & SMARTS \\
3241.31 & 1.24 & 1.481 & 0.004 & 1.846 & 0.007 & 2.293 & 0.007 & EULER \\
3258.72 & 1.36 & 1.493 & 0.008 & 1.844 & 0.023 & 2.310 & 0.042 & SMARTS \\
3270.73 & 1.87 & 1.490 & 0.035 & 1.838 & 0.088 & 2.341 & 0.111 & SMARTS \\
3279.64 & 1.53 & 1.502 & 0.011 & 1.794 & 0.045 & 2.270 & 0.034 & SMARTS \\
3282.63 & 1.25 & 1.494 & 0.011 & 1.821 & 0.025 & 2.296 & 0.035 & SMARTS \\
3287.60 & 1.66 & 1.505 & 0.016 & 1.840 & 0.019 & 2.300 & 0.037 & SMARTS \\
3291.56 & 1.60 & 1.497 & 0.025 & 1.800 & 0.050 & 2.290 & 0.066 & SMARTS \\
3292.57 & 1.89 & 1.489 & 0.023 & 1.872 & 0.050 & 2.349 & 0.061 & SMARTS \\
3295.62 & 1.10 & 1.483 & 0.016 & 1.835 & 0.045 & 2.297 & 0.060 & SMARTS \\
3296.15 & 1.50 & 1.512 & 0.008 & 1.857 & 0.022 & 2.302 & 0.023 & EULER \\
3298.57 & 1.20 & 1.508 & 0.014 & 1.840 & 0.016 & 2.312 & 0.023 & SMARTS \\
3301.62 & 1.65 & 1.464 & 0.036 & 1.831 & 0.094 & 2.329 & 0.136 & SMARTS \\
3302.09 & 1.48 & 1.486 & 0.011 & 1.854 & 0.033 & 2.309 & 0.051 & EULER \\
3303.06 & 1.45 & 1.496 & 0.015 & 1.870 & 0.030 & 2.330 & 0.035 & EULER \\
3303.58 & 1.71 & 1.477 & 0.018 & 1.837 & 0.052 & 2.324 & 0.069 & SMARTS \\
3307.59 & 1.62 & 1.444 & 0.016 & 1.776 & 0.047 & 2.223 & 0.078 & SMARTS \\
3309.08 & 1.24 & 1.492 & 0.002 & 1.873 & 0.012 & 2.309 & 0.010 & EULER \\
3310.13 & 1.03 & 1.473 & 0.007 & 1.863 & 0.019 & 2.293 & 0.022 & EULER \\
3310.55 & 1.22 & 1.473 & 0.033 & 1.874 & 0.101 & 2.319 & 0.112 & SMARTS \\
3320.55 & 1.27 & 1.499 & 0.013 & 1.891 & 0.041 & 2.317 & 0.037 & SMARTS \\
3324.56 & 1.32 & 1.485 & 0.010 & 1.870 & 0.030 & 2.294 & 0.046 & SMARTS \\
3328.52 & 1.49 & 1.485 & 0.019 & 1.834 & 0.045 & 2.267 & 0.047 & SMARTS \\
3329.07 & 1.05 & 1.499 & 0.011 & 1.857 & 0.031 & 2.269 & 0.033 & EULER \\
3338.51 & 1.64 & 1.492 & 0.021 & 1.760 & 0.047 & 2.254 & 0.080 & SMARTS \\
3347.06 & 1.18 & 1.518 & 0.006 & 1.847 & 0.016 & 2.293 & 0.019 & EULER \\
3432.39 & 1.30 & 1.475 & 0.010 & 1.824 & 0.028 & 2.266 & 0.025 & EULER \\
3434.40 & 1.20 & 1.477 & 0.006 & 1.809 & 0.017 & 2.270 & 0.029 & EULER \\
3442.37 & 1.92 & 1.465 & 0.015 & 1.854 & 0.020 & 2.300 & 0.024 & EULER \\
3450.39 & 1.08 & 1.481 & 0.005 & 1.850 & 0.010 & 2.265 & 0.012 & EULER \\
3458.40 & 1.29 & 1.477 & 0.013 & 1.823 & 0.040 & 2.257 & 0.043 & EULER \\
3480.38 & 1.58 & 1.493 & 0.015 & 1.871 & 0.040 & 2.302 & 0.042 & EULER \\
3483.86 & 1.27 & 1.491 & 0.022 & 1.871 & 0.028 & 2.266 & 0.027 & SMARTS \\
3500.40 & 1.62 & 1.482 & 0.008 & 1.942 & 0.022 & 2.305 & 0.029 & EULER \\
3508.85 & 1.53 & 1.496 & 0.030 & 1.939 & 0.066 & 2.311 & 0.072 & SMARTS \\
3511.34 & 1.41 & 1.533 & 0.005 & 1.983 & 0.020 & 2.320 & 0.022 & EULER \\
3516.36 & 1.18 & 1.549 & 0.011 & 2.019 & 0.026 & 2.301 & 0.019 & EULER \\
3520.41 & 1.40 & 1.545 & 0.012 & 1.965 & 0.025 & 2.293 & 0.025 & EULER \\
3520.87 & 1.13 & 1.556 & 0.014 & 1.955 & 0.036 & 2.249 & 0.030 & SMARTS \\
3522.41 & 1.00 & 1.545 & 0.006 & 1.970 & 0.014 & 2.287 & 0.018 & EULER \\
3524.38 & 1.34 & 1.560 & 0.029 & 2.024 & 0.074 & 2.361 & 0.073 & EULER \\
3528.83 & 1.16 & 1.558 & 0.010 & 1.948 & 0.016 & 2.314 & 0.018 & SMARTS \\
3547.35 & 1.26 & 1.596 & 0.006 & 1.928 & 0.019 & 2.336 & 0.021 & EULER \\
3558.23 & 1.38 & 1.587 & 0.005 & 1.887 & 0.019 & 2.377 & 0.015 & EULER \\
3562.86 & 1.37 & 1.611 & 0.051 & 1.877 & 0.142 & 2.371 & 0.173 & SMARTS \\
3585.76 & 1.22 & 1.580 & 0.013 & 1.825 & 0.035 & 2.409 & 0.052 & SMARTS \\
3586.27 & 1.61 & 1.546 & 0.005 & 1.804 & 0.008 & 2.395 & 0.015 & EULER \\
3592.04 & 1.05 & 1.537 & 0.011 & 1.822 & 0.024 & 2.384 & 0.029 & EULER \\
3601.19 & 0.97 & 1.512 & 0.009 & 1.788 & 0.024 & 2.362 & 0.029 & EULER \\
3602.21 & 1.42 & 1.497 & 0.006 & 1.760 & 0.030 & 2.333 & 0.030 & EULER \\
3603.27 & 1.24 & 1.507 & 0.008 & 1.784 & 0.024 & 2.387 & 0.027 & EULER \\
3606.21 & 1.91 & 1.490 & 0.013 & 1.757 & 0.035 & 2.363 & 0.039 & EULER \\
3606.72 & 1.75 & 1.487 & 0.016 & 1.763 & 0.041 & 2.355 & 0.068 & SMARTS \\
3607.18 & 1.22 & 1.477 & 0.008 & 1.766 & 0.018 & 2.395 & 0.020 & EULER \\
3608.17 & 1.34 & 1.475 & 0.006 & 1.760 & 0.012 & 2.366 & 0.019 & EULER \\
3614.32 & 1.41 & 1.461 & 0.015 & 1.771 & 0.036 & 2.403 & 0.043 & EULER \\
3614.70 & 1.07 & 1.476 & 0.003 & 1.761 & 0.023 & 2.383 & 0.009 & SMARTS \\
3620.69 & 1.27 & 1.475 & 0.004 & 1.753 & 0.012 & 2.346 & 0.021 & SMARTS \\
3633.64 & 1.58 & 1.460 & 0.032 & 1.721 & 0.054 & 2.311 & 0.063 & SMARTS \\
3638.66 & 1.64 & 1.454 & 0.013 & 1.736 & 0.034 & 2.359 & 0.052 & SMARTS \\
3640.24 & 1.47 & 1.454 & 0.039 & 1.747 & 0.098 & 2.331 & 0.129 & EULER \\
3641.58 & 1.53 & 1.450 & 0.022 & 1.737 & 0.057 & 2.359 & 0.077 & SMARTS \\
3643.61 & 1.57 & 1.434 & 0.011 & 1.768 & 0.014 & 2.315 & 0.069 & SMARTS \\
3648.60 & 0.96 & 1.428 & 0.048 & 1.686 & 0.118 & 2.348 & 0.218 & SMARTS \\
3650.11 & 1.18 & 1.421 & 0.006 & 1.709 & 0.015 & 2.298 & 0.016 & EULER \\
3651.55 & 1.05 & 1.412 & 0.027 & 1.678 & 0.064 & 2.293 & 0.085 & SMARTS \\
3654.58 & 1.60 & 1.417 & 0.015 & 1.687 & 0.040 & 2.326 & 0.055 & SMARTS \\
3665.55 & 1.29 & 1.420 & 0.020 & 1.729 & 0.025 & 2.289 & 0.042 & SMARTS \\
3668.06 & 1.43 & 1.433 & 0.010 & 1.722 & 0.020 & 2.271 & 0.023 & EULER \\
3668.52 & 1.50 & 1.415 & 0.014 & 1.690 & 0.031 & 2.276 & 0.062 & SMARTS \\
3672.10 & 1.43 & 1.432 & 0.007 & 1.732 & 0.014 & 2.302 & 0.018 & EULER \\
3673.52 & 1.53 & 1.411 & 0.014 & 1.669 & 0.039 & 2.253 & 0.053 & SMARTS \\
3675.05 & 1.19 & 1.432 & 0.009 & 1.745 & 0.017 & 2.296 & 0.026 & EULER \\
3675.52 & 1.12 & 1.395 & 0.011 & 1.686 & 0.034 & 2.271 & 0.042 & SMARTS \\
3676.07 & 1.53 & 1.406 & 0.014 & 1.707 & 0.038 & 2.293 & 0.046 & EULER \\
3677.11 & 1.50 & 1.408 & 0.011 & 1.697 & 0.024 & 2.281 & 0.033 & EULER \\
3678.07 & 1.46 & 1.402 & 0.012 & 1.716 & 0.026 & 2.283 & 0.027 & EULER \\
3680.07 & 1.43 & 1.394 & 0.007 & 1.712 & 0.023 & 2.290 & 0.034 & EULER \\
3680.51 & 1.49 & 1.388 & 0.021 & 1.701 & 0.043 & 2.240 & 0.051 & SMARTS \\
3681.10 & 1.78 & 1.395 & 0.018 & 1.706 & 0.037 & 2.289 & 0.032 & EULER \\
3682.12 & 1.86 & 1.387 & 0.023 & 1.729 & 0.052 & 2.303 & 0.056 & EULER \\
3682.51 & 1.70 & 1.389 & 0.023 & 1.673 & 0.064 & 2.218 & 0.069 & SMARTS \\
3684.07 & 1.63 & 1.365 & 0.013 & 1.694 & 0.031 & 2.273 & 0.042 & EULER \\
3685.10 & 1.22 & 1.391 & 0.010 & 1.710 & 0.028 & 2.292 & 0.032 & EULER \\
3686.07 & 1.44 & 1.380 & 0.008 & 1.695 & 0.024 & 2.305 & 0.032 & EULER \\
3687.07 & 1.45 & 1.362 & 0.011 & 1.682 & 0.026 & 2.286 & 0.029 & EULER \\
3687.51 & 1.50 & 1.375 & 0.009 & 1.705 & 0.031 & 2.262 & 0.083 & SMARTS \\
3688.10 & 1.10 & 1.386 & 0.012 & 1.733 & 0.030 & 2.270 & 0.029 & EULER \\
3689.06 & 1.33 & 1.379 & 0.018 & 1.713 & 0.031 & 2.305 & 0.037 & EULER \\
3690.08 & 1.65 & 1.380 & 0.021 & 1.711 & 0.056 & 2.273 & 0.100 & EULER \\
3691.05 & 1.65 & 1.383 & 0.029 & 1.719 & 0.072 & 2.302 & 0.079 & EULER \\
3692.03 & 1.64 & 1.382 & 0.011 & 1.762 & 0.031 & 2.279 & 0.041 & EULER \\
3693.06 & 1.17 & 1.378 & 0.007 & 1.729 & 0.020 & 2.268 & 0.021 & EULER \\
3694.06 & 1.22 & 1.378 & 0.004 & 1.730 & 0.014 & 2.256 & 0.012 & EULER \\
3695.06 & 1.72 & 1.370 & 0.008 & 1.724 & 0.021 & 2.297 & 0.021 & EULER \\
3696.07 & 1.55 & 1.367 & 0.013 & 1.742 & 0.020 & 2.230 & 0.026 & EULER \\
3699.52 & 1.30 & 1.370 & 0.017 & 1.727 & 0.048 & 2.246 & 0.057 & SMARTS \\
3700.07 & 1.60 & 1.376 & 0.012 & 1.762 & 0.028 & 2.246 & 0.032 & EULER \\
3701.53 & 1.58 & 1.365 & 0.006 & 1.692 & 0.066 & 2.183 & 0.047 & SMARTS \\
3707.07 & 1.79 & 1.394 & 0.015 & 1.763 & 0.035 & 2.253 & 0.048 & EULER \\
3806.39 & 1.42 & 1.370 & 0.010 & 1.717 & 0.023 & 2.234 & 0.030 & EULER \\
3819.38 & 1.19 & 1.353 & 0.010 & 1.686 & 0.026 & 2.225 & 0.037 & EULER \\
3820.34 & 1.28 & 1.377 & 0.011 & 1.710 & 0.024 & 2.222 & 0.027 & EULER \\
3824.39 & 1.32 & 1.384 & 0.004 & 1.712 & 0.012 & 2.219 & 0.013 & EULER \\
3826.89 & 1.25 & 1.371 & 0.017 & 1.679 & 0.042 & 2.218 & 0.046 & SMARTS \\
3829.37 & 1.24 & 1.358 & 0.007 & 1.697 & 0.017 & 2.177 & 0.020 & EULER \\
3831.39 & 1.09 & 1.366 & 0.006 & 1.720 & 0.020 & 2.185 & 0.019 & EULER \\
3832.41 & 1.18 & 1.370 & 0.005 & 1.723 & 0.011 & 2.178 & 0.013 & EULER \\
3839.87 & 1.79 & 1.393 & 0.011 & 1.714 & 0.026 & 2.153 & 0.069 & SMARTS \\
3844.41 & 1.21 & 1.353 & 0.011 & 1.739 & 0.030 & 2.216 & 0.031 & EULER \\
3847.40 & 1.37 & 1.352 & 0.005 & 1.732 & 0.010 & 2.188 & 0.011 & EULER \\
3848.39 & 1.54 & 1.350 & 0.009 & 1.725 & 0.020 & 2.204 & 0.025 & EULER \\
3849.39 & 1.16 & 1.351 & 0.005 & 1.746 & 0.015 & 2.199 & 0.017 & EULER \\
3850.37 & 1.46 & 1.354 & 0.008 & 1.744 & 0.032 & 2.200 & 0.027 & EULER \\
3851.34 & 1.28 & 1.339 & 0.007 & 1.757 & 0.020 & 2.187 & 0.025 & EULER \\
3852.35 & 1.18 & 1.352 & 0.006 & 1.767 & 0.018 & 2.184 & 0.018 & EULER \\
3852.87 & 1.19 & 1.383 & 0.007 & 1.787 & 0.019 & 2.230 & 0.027 & SMARTS \\
3863.79 & 1.07 & 1.371 & 0.011 & 1.809 & 0.029 & 2.181 & 0.038 & SMARTS \\
3869.36 & 1.23 & 1.368 & 0.015 & 1.824 & 0.036 & 2.168 & 0.039 & EULER \\
3877.88 & 1.15 & 1.390 & 0.010 & 1.858 & 0.037 & 2.219 & 0.028 & SMARTS \\
3886.86 & 0.84 & 1.422 & 0.015 & 1.893 & 0.040 & 2.207 & 0.043 & SMARTS \\
3887.42 & 1.29 & 1.406 & 0.006 & 1.861 & 0.014 & 2.199 & 0.023 & EULER \\
3889.40 & 1.12 & 1.415 & 0.005 & 1.884 & 0.017 & 2.204 & 0.017 & EULER \\
3891.42 & 1.32 & 1.424 & 0.005 & 1.881 & 0.017 & 2.199 & 0.014 & EULER \\
3892.40 & 1.19 & 1.412 & 0.009 & 1.841 & 0.043 & 2.177 & 0.029 & EULER \\
3896.88 & 1.72 & 1.463 & 0.024 & 1.892 & 0.078 & 2.236 & 0.116 & SMARTS \\
3900.39 & 1.09 & 1.430 & 0.006 & 1.883 & 0.020 & 2.187 & 0.016 & EULER \\
3903.79 & 1.23 & 1.460 & 0.015 & 1.922 & 0.053 & 2.217 & 0.067 & SMARTS \\
3908.33 & 1.10 & 1.455 & 0.008 & 1.882 & 0.015 & 2.202 & 0.018 & EULER \\
3913.28 & 1.08 & 1.476 & 0.007 & 1.891 & 0.017 & 2.222 & 0.015 & EULER \\
3917.12 & 1.72 & 1.477 & 0.010 & 1.841 & 0.026 & 2.244 & 0.023 & EULER \\
3918.84 & 0.60 & 1.521 & 0.020 & 1.851 & 0.048 & 2.182 & 0.091 & SMARTS \\
3919.33 & 1.11 & 1.474 & 0.006 & 1.860 & 0.034 & 2.209 & 0.013 & EULER \\
3932.40 & 1.53 & 1.526 & 0.020 & 1.812 & 0.049 & 2.297 & 0.061 & EULER \\
3937.73 & 1.18 & 1.537 & 0.008 & 1.802 & 0.015 & 2.301 & 0.006 & SMARTS \\
3944.22 & 1.63 & 1.522 & 0.022 & 1.797 & 0.040 & 2.343 & 0.073 & EULER \\
3944.78 & 1.58 & 1.544 & 0.005 & 1.765 & 0.017 & 2.330 & 0.022 & SMARTS \\
3945.35 & 1.35 & 1.510 & 0.007 & 1.764 & 0.015 & 2.269 & 0.016 & EULER \\
3946.32 & 1.33 & 1.517 & 0.007 & 1.758 & 0.015 & 2.291 & 0.017 & EULER \\
3946.63 & 1.39 & 1.531 & 0.010 & 1.768 & 0.030 & 2.278 & 0.025 & SMARTS \\
3950.27 & 1.35 & 1.489 & 0.006 & 1.737 & 0.017 & 2.296 & 0.025 & EULER \\
3952.24 & 1.20 & 1.488 & 0.047 & 1.780 & 0.102 & 2.377 & 0.125 & EULER \\
3961.31 & 1.57 & 1.463 & 0.009 & 1.755 & 0.019 & 2.342 & 0.026 & EULER \\
3964.09 & 1.11 & 1.449 & 0.004 & 1.752 & 0.011 & 2.322 & 0.018 & EULER \\
3964.75 & 0.74 & 1.474 & 0.021 & 1.719 & 0.055 & 2.331 & 0.070 & SMARTS \\
3967.78 & 0.67 & 1.450 & 0.011 & 1.738 & 0.029 & 2.340 & 0.033 & SMARTS \\
3968.76 & 1.24 & 1.448 & 0.020 & 1.761 & 0.038 & 2.317 & 0.048 & SMARTS \\
3970.22 & 1.48 & 1.447 & 0.011 & 1.778 & 0.037 & 2.293 & 0.041 & EULER \\
3971.75 & 1.02 & 1.453 & 0.013 & 1.776 & 0.029 & 2.325 & 0.036 & SMARTS \\
3979.17 & 1.35 & 1.426 & 0.003 & 1.778 & 0.013 & 2.313 & 0.017 & EULER \\
3979.75 & 1.66 & 1.452 & 0.018 & 1.840 & 0.031 & 2.248 & 0.050 & SMARTS \\
3980.27 & 1.56 & 1.435 & 0.018 & 1.774 & 0.031 & 2.291 & 0.035 & EULER \\
3981.25 & 1.10 & 1.415 & 0.010 & 1.786 & 0.024 & 2.293 & 0.036 & EULER \\
3982.25 & 1.63 & 1.417 & 0.013 & 1.767 & 0.031 & 2.303 & 0.048 & EULER \\
3987.55 & 1.34 & 1.421 & 0.021 & 1.818 & 0.068 & 2.351 & 0.083 & SMARTS \\
3994.14 & 1.38 & 1.396 & 0.007 & 1.805 & 0.024 & 2.265 & 0.025 & EULER \\
3994.64 & 1.03 & 1.400 & 0.017 & 1.776 & 0.022 & 2.272 & 0.021 & SMARTS \\
3998.12 & 1.88 & 1.404 & 0.010 & 1.798 & 0.025 & 2.258 & 0.068 & EULER \\
3999.11 & 1.22 & 1.401 & 0.008 & 1.800 & 0.015 & 2.263 & 0.013 & EULER \\
4002.57 & 0.91 & 1.416 & 0.010 & 1.837 & 0.024 & 2.279 & 0.035 & SMARTS \\
4003.10 & 1.00 & 1.405 & 0.005 & 1.817 & 0.015 & 2.271 & 0.013 & EULER \\
4005.10 & 1.37 & 1.414 & 0.011 & 1.831 & 0.032 & 2.248 & 0.031 & EULER \\
4007.62 & 1.26 & 1.429 & 0.011 & 1.834 & 0.011 & 2.273 & 0.017 & SMARTS \\
4008.17 & 1.18 & 1.407 & 0.007 & 1.820 & 0.016 & 2.248 & 0.022 & EULER \\
4014.53 & 1.43 & 1.440 & 0.014 & 1.829 & 0.067 & 2.241 & 0.069 & SMARTS \\
4021.58 & 1.12 & 1.439 & 0.011 & 1.894 & 0.038 & 2.256 & 0.038 & SMARTS \\
4024.15 & 1.60 & 1.435 & 0.009 & 1.852 & 0.020 & 2.272 & 0.023 & EULER \\
4025.10 & 1.72 & 1.442 & 0.007 & 1.882 & 0.022 & 2.266 & 0.025 & EULER \\
4027.09 & 1.84 & 1.463 & 0.008 & 1.876 & 0.030 & 2.254 & 0.029 & EULER \\
4028.57 & 1.09 & 1.456 & 0.006 & 1.844 & 0.024 & 2.295 & 0.019 & SMARTS \\
4032.06 & 1.81 & 1.446 & 0.015 & 1.850 & 0.037 & 2.271 & 0.046 & EULER \\
4036.07 & 1.45 & 1.452 & 0.007 & 1.849 & 0.022 & 2.258 & 0.022 & EULER \\
4036.53 & 1.07 & 1.478 & 0.014 & 1.828 & 0.026 & 2.295 & 0.016 & SMARTS \\
4039.05 & 1.67 & 1.470 & 0.012 & 1.877 & 0.030 & 2.237 & 0.034 & EULER \\
4042.04 & 1.34 & 1.475 & 0.008 & 1.839 & 0.027 & 2.283 & 0.025 & EULER \\
4046.04 & 1.52 & 1.492 & 0.007 & 1.830 & 0.023 & 2.283 & 0.023 & EULER \\
4057.05 & 1.90 & 1.507 & 0.010 & 1.776 & 0.034 & 2.261 & 0.034 & EULER \\
4065.06 & 1.29 & 1.499 & 0.006 & 1.774 & 0.022 & 2.313 & 0.027 & EULER \\
4072.06 & 1.87 & 1.514 & 0.011 & 1.774 & 0.023 & 2.359 & 0.064 & EULER \\
4174.40 & 1.58 & 1.436 & 0.022 & 1.730 & 0.058 & 2.277 & 0.080 & EULER \\
4183.40 & 1.43 & 1.397 & 0.010 & 1.709 & 0.031 & 2.279 & 0.037 & EULER \\
4191.40 & 1.61 & 1.404 & 0.015 & 1.737 & 0.026 & 2.226 & 0.036 & EULER \\
4197.36 & 1.42 & 1.402 & 0.017 & 1.740 & 0.035 & 2.230 & 0.042 & EULER \\
4203.39 & 1.13 & 1.400 & 0.011 & 1.734 & 0.025 & 2.223 & 0.031 & EULER \\
4204.39 & 1.43 & 1.400 & 0.011 & 1.756 & 0.030 & 2.256 & 0.034 & EULER \\
4207.37 & 1.60 & 1.411 & 0.009 & 1.739 & 0.031 & 2.236 & 0.034 & EULER \\
4213.41 & 1.63 & 1.404 & 0.007 & 1.748 & 0.019 & 2.229 & 0.027 & EULER \\
4228.34 & 1.71 & 1.397 & 0.020 & 1.697 & 0.047 & 2.198 & 0.055 & EULER \\
4230.31 & 1.22 & 1.392 & 0.008 & 1.697 & 0.024 & 2.240 & 0.025 & EULER \\
\hline
\end{longtable}
}
\end{document}